\documentclass{aa}  
\usepackage[normalem]{ulem}
\usepackage{placeins}
\usepackage{color}
\usepackage{calc}
\usepackage{amsmath,amssymb,graphicx}
\usepackage{tensor}
\usepackage{bm}
\usepackage{times}
\usepackage[varg]{txfonts}
\usepackage{float}
\usepackage{dcolumn}
\usepackage[nolist,nohyperlinks]{acronym}
\usepackage{xspace}
\usepackage[abs]{overpic}
\usepackage{pict2e}
\usepackage{enumitem}
\usepackage[usenames,dvipsnames]{xcolor}
\usepackage[utf8]{inputenc}
\usepackage{acronym}
\usepackage{gensymb}
\usepackage{caption}

\usepackage[normalem]{ulem}
\usepackage{longtable}
\usepackage{verbatim}
\usepackage{multirow}
\usepackage{amsmath,amssymb}
\usepackage{csquotes}
\usepackage[export]{adjustbox}

\usepackage{subfigure}
\usepackage{placeins}
\usepackage[pdfborder={0 0 0},colorlinks=true,citecolor=NavyBlue]{hyperref}
\usepackage{linenoaa}
\usepackage{soul}

\usepackage{cleveref}

\usepackage{hhline}

\newcommand{\Msun}{\ensuremath{M_{ \odot }}}

\begin{document}

\title{Reconciling PTA and JWST and preparing for LISA with \texttt{POMPOCO}: Parametrisation Of the Massive black hole POpulation for Comparison to Observations}

\author{A. Toubiana \inst{1,2,3} \and L. Sberna \inst{4} \and M. Volonteri \inst{5} \and E. Barausse \inst{6,7}  \and S. Babak \inst{8}   \and R. Enficiaud \inst{1} \and D. Izquierdo–Villalba \inst{2,3}  \and J. R. Gair \inst{1}  \and J. E. Greene\inst{9} \and  H. Quelquejay Leclere \inst{8} }

\institute{Max Planck Institute for Gravitationsphysik (Albert Einstein Institute), Am M\"{u}hlenberg 1, 14476 Potsdam, Germany 
\and
Dipartimento di Fisica “G. Occhialini”, Universit´a degli Studi di Milano-Bicocca, Piazza della Scienza 3, 20126 Milano, Italy
\and 
INFN, Sezione di Milano-Bicocca, Piazza della Scienza 3, 20126 Milano, Italy
\and 
School of Mathematical Sciences, University of Nottingham, University Park, Nottingham NG7 2RD, United Kingdom 
\and
  Institut d’Astrophysique de Paris, UMR 7095, CNRS and Sorbonne Université, 98 bis boulevard Arago, 75014 Paris, France 
\and 
SISSA - Scuola Internazionale Superiore di Studi Avanzati, Via Bonomea 265, 34136 Trieste, Italy and INFN Sezione di Trieste 
\and 
IFPU - Institute for Fundamental Physics of the Universe, Via Beirut 2, 34014 Trieste, Italy 
\and 
Université Paris Cité, CNRS, Astroparticule et Cosmologie, F-75013 Paris, France
\and
Department of Astrophysical Sciences, Princeton University, Princeton, NJ 08544, USA
}

 \abstract
   {}
   { We develop a parametrised model to describe the formation and evolution of massive black holes, designed for comparisons with both electromagnetic and gravitational wave observations.}
  {
  Using an extended Press-Schechter formalism, we generated tter halo merger trees. We then seeded and evolved massive black holes through parametrised prescriptions. This approach avoids solving differential equations and is computationally efficient. It enabled us to analyse observational data and infer the parameters of our model in a fully Bayesian framework.
  }
   {Observations of the black hole luminosity function are compatible with the nHz gravitational wave signal (likely) measured by pulsar timing arrays (PTAs), provided we allow for an increased luminosity function at high redshift ($4-7$), as recently suggested by JWST observations. 
   Our model can simultaneously reproduce the bulk of the $M_*-M_{\rm BH}$ relation at $z-0$, as well as its outliers, something cosmological simulations struggle to do.
   The inferred model parameters are consistent with expectations from observations and more complex simulations: They favour heavier black hole seeds and short delays between halo and black hole mergers, while requiring supper-Edington accretion episodes lasting a few tens of million years, which in our model are linked to galaxy mergers. Accretion is suppressed in the most massive black holes below $z\simeq 2.5$ in our model, which is consistent with the anti-hierarchical growth hypothesis. Finally, our predictions for LISA, although fairly broad, agree with previous models that assumed an efficient merging of massive black holes formed from heavy seeds.} 
   {Our model offers a new perspective on the apparent tensions between the black hole luminosity function and the latest JWST and PTA results. Its flexibility makes it ideal to fully exploit the potential of future gravitational-wave observations of massive black hole binaries with LISA.  }


\maketitle

\section{Introduction}

The general scenario is as follows. MBHs must have originated at high redshift from lighter black hole ``seeds''. Several seed models have been proposed, ranging from ``light'' to ``heavy'' \citep[for a review, see][]{Latif:2016qau,Volonteri2021review}. At the largest scales, the fate of a seed is determined by the evolution of its host dark matter (DM) halo, which proceeds hierarchically in mass. Within the host galaxy, the growth of the seed proceeds primarily through the accretion of gas~\citep{Yu_2002}, and in a minor part via the accretion of stars \citep{Rees:1988bf}, with periods of intense accretion activity likely alternating with quiescent phases. Intense accretion emits significant energy in a variety of forms: Radiation is emitted across the electromagnetic spectrum \citep{Fabian2012}, while kinetic and thermal energy can affect the gas temperature, density, and turbulence, and it can in turn prevent new gas from reaching the MBH and temporarily suppress the process (AGN feedback). To a lesser extent \citep{2019MNRAS.489..802R}, BH mergers following the merger of their host galaxies can also contribute to the growth of MBHs \citep{Dubois:2013rha,Kulier:2013gda}.

One of our best probes of the population of MBHs is the luminosity function (LF) of accreting MBHs: quasars and active galactic nuclei (AGNs) more generally. This has been the target of surveys in the X-ray, UV, optical, infrared, and radio bands \citep[see][and references therein]{Hopkins:2006fq,Shen:2020obl}. These surveys, covering out to $z\lesssim 7$, have revealed that the LF evolves strongly with redshift in normalisation and shape . 
Interestingly, mid-infrared surveys appeared in tension with the aforementioned results. Using observations from the Spitzer Space Telescope survey \citep{2004ApJS..154....1W}, \cite{Lacy:2015tha} found an increased LF up to $z\sim3$. The higher-redshift mid-infrared Universe is now being probed by the James Webb Space Telescope (JWST) \citep{Gardner:2006ky}. It has detected tens of candidate AGNs at high redshift ($z \sim 4 \, \textrm{--} \, 11$), including some at lower bolometric luminosities than were observed before ($L_{\rm bol} \sim 10^{44} \, \textrm{--} \, 10^{46} {\rm erg}/{\rm s}$) 
\citep[e.g.,][]{Onoue2023,Kocevski2023,Maiolino:2023bpi,2023A&A...677A.145U,Kokorev2023,Lyu__2024}. The first results for the LF and mass function derived from these observations \citep{harikane2023jwstnirspeccensusbroadlineagns,Maiolino:2023bpi,Greene2024,Matthee2024,2025ApJ...986..165T} also point to an increased LF compared to previous expectations -- based on the compilation of data from mid-IR to X-ray, but dominated at high redshift by X-ray and UV observations \citep{Shen:2020obl}. It has been suggested that the masses and abundance of (candidate) MBHs found by JWST at very high redshifts might pose a challenge for lighter seed scenarios, and might require continuous and intense accretion even for heavier seeds \citep[e.g.,][]{harikane2023jwstnirspeccensusbroadlineagns,Maiolino:2023bpi,Kokorev2023,Greene2024}. At the same time, the models that produce the heaviest seeds appear to struggle to produce enough seeds to explain the abundance of (candidate) MBHs \citep{Regan2024} (see also \citet{Habouzit2024} for a discussion of the compatibility of different simulations of MBH evolution with these results).  

These electromagnetic observations are now being complemented with gravitational-wave (GW) observations, as pulsar timing array (PTA) collaborations are likely close to confirming the detection of a stochastic background consistent with merging MBH binaries with masses $\gtrsim 10^8 M_{\odot}$ \citep{EPTA:2023fyk,NANOGrav:2023gor,Tarafdar:2022toa,Reardon:2023gzh,Xu:2023wog}. It is generally thought that the preliminary result for the PTA background, with its relatively high amplitude, would require MBHs to merge and accrete efficiently \citep[]{NANOGrav:2023hfp,EPTA:2023xxk,Barausse:2023yrx,2024A&A...686A.183I} (although see \cite{Goncharov:2024htb}). In the next decade, the Laser Interferometer Space Antenna (LISA) will further extend our understanding of merging MBHs by detecting GWs from binaries with lower masses ($10^4 \textrm{--} 10^8 M_\odot$) and up to very high redshifts ($z\sim 20$) \citep{amaroseoane2017laserinterferometerspaceantenna,2024arXiv240207571C}.

The astrophysical interpretation of observations is achieved by a comparison to theoretical models for the formation and evolution of MBHs. 
Cosmological $N-{\rm body}$ and hydrodynamic simulations have become valuable tools for simulating MBHs, thanks to increased computational power and more accurate sub-grid models for MBH accretion, seeding, feedback, and dynamics: see, e.g.~\citet{Volonteri_2016,Springel_2017, Kannan_2021,Ni_2022,Bhowmick2024}. In addition, semi-analytical models \citep[SAMs; e.g.,][]{Cole_2002,Volonteri:2002vz,Monaco2007,Somerville_2008,Benson:2010kx,Barausse:2012fy,Ricarte:2017ihq,Bonetti:2018tpf,Dayal_2019,Izquierdo-Villalba:2020hfk,Barausse:2020mdt,Trinca_2022} have been widely used to simulate the cosmic history of MBHs thanks to their reduced computational cost. A computationally cheaper alternative to both SAMs and cosmological simulations are empirical models \citep[see, e.g.,][]{Soltan1982,Small1992,Tucci:2016tyc,Conroy_2012,Allevato_2021}. Instead of making physical assumptions, these models use observations to empirically (and self-consistently) characterise the evolution of MBHs. Empirical models have grown increasingly more complex, and some \citep{Zhang2023,Boettner:2023rdt} encompass DM halos, galaxies and MBHs and count about 50 parameters. The more complex the model, the greater the number of observations that must be used to constrain its parameters, from scaling relations to luminosity and mass functions.

Cosmological simulations and SAMs are too computationally expensive to compare the full range of alternative models and parameter space with observations. This results in a set of discrete models. This is a major obstacle for astrophysical inference within a Bayesian framework. In \cite{Toubiana:2021iuw}, some of the authors of this paper studied mock LISA data and showed that inference with a finite set of discrete population models could lead to severe biases in the underlying astrophysical parameters. Empirical models, while sufficiently fast for a Bayesian parameter estimation \citep[see, e.g.,][]{Zhang2023,Boettner:2023rdt}, focus on statistical and empirical properties, and not on the underlying physics.

The goal of this work is to provide a \emph{fast} parametric approach to infer the physics behind the evolution of MBHs from current observations, and prepare for the LISA mission. We introduce our model \texttt{POMPOCO}: {\it Parametrisation Of the Massive black hole POpulation for Comparison to Observations}. It adopts an intermediate approach between SAMs and empirical models, by proposing an effective description of the formation and evolution of MBHs within their host haloes. 
As a first application of \texttt{POMPOCO}, we jointly fit the LF at low and high redshifts and the amplitude of the GW background measured by PTA, using a Markov-Chain-Monte-Carlo (MCMC) algorithm to find the sets of parameters that are compatible with the desired datasets. 
We validate our results on the $M_*-M_{\rm BH}$ relation at $z=0$ and at higher redshift, for which we do not explicitly fit. The results agree well with the proposed fit of \cite{Greene2020}. 
Finally, we compute predictions for LISA conditioned on these observations.

This paper is organised as follows. Section~\ref{sec:model} describes our parametric model. In Sec.~\ref{sec:observables} and \ref{sec:fits}, we construct observables from our model, and use them to fit the model parameters to observations. We present our results in Sec.~\ref{sec:results}, and make predictions for the LISA mission in Sec.~\ref{sec:lisa}. We summarise and conclude in Sec.~\ref{sec:conclusions}. In the remainder of the paper, MBH masses are defined in the source frame.

\section{Description of the model}\label{sec:model}
We describe the components of our model below. We also summarise all the parameters and the settings of our model in Table~\ref{tab:params}. In general, our choice of prescriptions is guided by a combination of physical, observational, and simulation-based considerations, as well as a deliberate effort to maintain flexibility and allow the data to inform the inference. In the current version of our model, we do not track the evolution of the spins of MBHs. We hope to do so in future work.

We adopt a flat $\Lambda$CDM cosmology with ${H_0=67.77 \ {\rm km} \; {\rm s}^{-1} {\rm Mpc}^{-1}}$, $\Omega_m=0.3071$, $\Omega_b=0.04820$ and  $\delta_{c,0}=1.686$ for the critical overdensity for spherical collapse at $z=0$. Given the tight constraints on cosmological parameters, we expect the uncertainty from cosmology to be small compared to the astrophysical uncertainty.

\subsection{Dark matter halo merger tree}

We generated DM halo merger trees down to $z_{\rm max}=20$ using the implementation of the extended Press-Schechter formalism \citep{Press:1973iz,Bond:1990iw,Bower:1991kf,Lacey:1993iv} described in App.~A of \cite{Parkinson:2007yh}. We used updated values given in \cite{Benson:2016pht} for the phenomenological modifications to the original extended Press-Schechter formalism, introduced to better reproduce cosmological simulations. We used a time-varying mass resolution, as proposed in \cite{Volonteri:2002vz}: $M_{h, {\rm res}}=10^{-3}M_{h,0}(1+z)^{-3.5}$, where $M_{h,0}$ is the mass of the halo at $z=0$. 

The suite of trees was built drawing the halo masses at $z=0$ from a log-uniform distribution. Each tree entered the calculation of population properties by reweighting the BHs in that tree by their Press-Schechter weight.

\subsection{Massive black hole seeds}

We populated leaf haloes -- the last haloes along the branches of the merger trees\footnote{Leaf haloes can be at $z<z_{\rm max}$, if their mass drops below the mass resolution.} -- with seed BHs. We only seeded haloes at $z\geq 10$ and with mass above a threshold $M_{h,{\rm seed}}$, with probability $f_{\rm seed}$.

The nature and mass spectrum of the original seeds of the observed population of MBHs is currently unconstrained. Proposed seeding mechanisms range from light seeds (below $10^3M_\odot$), resulting from the first stars or formed primordially, and heavy seeds (between $10^3M_\odot$ and $10^6M_\odot$) resulting from runaway accretion, mergers or supermassive stars. For a recent review of these mechanisms, which are not mutually exclusive, see \cite{Volonteri2021review}. 
We allowed our model to probe the seed distribution by drawing the mass of the seed BH from a truncated log-normal distribution, with mean $\mu_{\rm seed}$ and standard deviation $\sigma_{\rm seed}$. The log-normal distribution is meant to capture a single formation scenario, characterised by a range of seed masses around a central value. At the high end, we limited the mass of the BH seed to $10\%$ of the baryonic mass of the halo $M_h\Omega_b/(\Omega_m-\Omega_b)$. At the low end, we took the minimum seed mass to be $100 \, M_{\odot}$.\footnote{We exclude stellar mass seeds below $m<100 \,M_{\odot}$, as~\citet{2018MNRAS.480.3762S}, \citet{2024A&A...691A..24S} found that less than $1\%$ can grow to contribute to the MBH population.}
We denoted the seeded BH as the \emph{primary} BH of the DM halo.

\subsection{Halo and massive black hole mergers}\label{sec:mbhb_merger}

Following the merger of two haloes, we tracked the evolution of the BHs they contain, distinguishing between \emph{major} and \emph{minor} halo mergers, based on the halo mass ratio $q_h=M_{h,2}/M_{h,1}\leq1$. Based on the results of our simulations, we set the threshold for major mergers to $q_{h,\rm major}= 0.13$, which is similar to the value used in \cite{Ricarte:2017ihq}. 

\subsubsection{Formation of black hole binaries}

Following major mergers, where $q_h \geq q_{h,\rm major}$, we checked whether either halo hosts a BH. If there was only one BH, this became the primary BH of the newly formed halo. If there was one BH in each, they formed a binary with merger time given by the sum of the dynamical friction (DF) timescale and additional delays (see Sec.~\ref{sec:delays}). Finally, if there were more BHs, we use the prescription for multiple interactions described below. 

 In the case of minor mergers, BHs in the lighter halo had to sink into the heavier halo before being counted as primary BH or forming a binary. The BHs of the lighter halo were therefore dubbed as \emph{outer} BHs of the heavier halo. The sinking time is given by the DF timescale of the satellite halo into the primary one. If the lighter halo contained a binary that merges before the sinking time, we allowed it to merge and assign the merger product to the new halo as an outer BH. We retained (the most massive) half of the outer BHs of the lighter halo, in addition to its potential primary/binary, up to a limit of four outer BHs. We find that this is an acceptable limitation, as small BHs in small haloes play a minor role. Similarly, following major mergers, we retained up to four of the outer BHs of the two haloes, selecting the ones with the shorter sinking times. When outer BHs sunk into the centre of the halo, they either became the primary BH of that halo if it contained none, or formed a binary that merged within $t_{\rm delay}$ (see Sec.~\ref{sec:delays}). If the halo already contained a binary, or multiple outer BHs sunk within one time step, we used our prescription for multiple interactions described in Sec.~\ref{sec:multi}.

\subsubsection{Multiple interactions}\label{sec:multi}

We used the results of \cite{Bonetti:2017dan} to handle interactions between more than two BHs. \cite{Bonetti:2017dan} explored the outcome of three-body interactions between MBHs 
accounting for relativistic corrections to the Newtonian equations of motion. They found that, on average, there is a probability $p_{\rm multi}=0.22$ (see their Table 2) that the interaction of an MBH binary with a third MBH will lead to a rapid merger (within a few hundred million years), usually between the two most massive MBHs. In the remaining cases, the MBH binary is little affected by the third MBH. 
Based on these results, we decided with probability $p_{\rm multi}=0.22$ if the BHs undergo a multiple interaction or not.  If it occurred, we retained the two most massive BHs and drew the merger time (in yr) from a log-normal distribution with mean $8.4 \  {\rm dex}$ and standard deviation $0.4 \ {\rm dex}$ (see Fig. 7 of \cite{Bonetti:2017dan}). If they did not undergo a multiple interaction, we distinguished between two cases, depending on how the BHs ``met''. If there was a binary and a third BH was brought either by a halo merger or by an outer BH that sunk, we kept the original binary, and its time to merger was unaffected. If there was no binary originally, for instance, if there was a primary BH and two outer BHs sunk within one time step, we just kept the two most massive among them, and formed a binary merging within $t_{\rm delay}$ (see Sec.~\ref{sec:delays}). In any case, we did not keep track of the other BHs.

\subsubsection{Dynamical friction and other delays}\label{sec:delays}

Satellite haloes (and their BHs) sink to the centre of the primary halo via DF, whose typical timescale is set by \citep{Lacey:1993iv,1987gady.book.....B}
\begin{align}
    t_{\rm DF}&=0.495 \,  \frac{1+q_h}{q_h}\frac{1}{H(z)\sqrt{\Delta_{\rm vir}}\ln(1+q_h)}, \\
    \Delta_{\rm vir}&=178 \, \Omega_m(z)^{0.45},\\
    q_h&=\frac{M_{h,2}}{M_{h,1}}, 
\end{align}
where we made the same choice for the numerical prefactor as in \cite{Volonteri:2002vz},\footnote{This numerical prefactor is related to the orbital parameters of the haloes. The value adopted in~\cite{Volonteri:2002vz} was motivated by numerical investigations.} and 
where $H(z)$ is the Hubble factor at the desired redshift. We neglected any possible variations in the DF timescale due to the initial conditions of the infall, and effects such as tidal stripping and tidal heating of the satellite halo.

Once the satellite halo has sunk into the primary one, the two galaxies they host will merge thanks to additional DF. The BHs at their centre will then form a bound binary, with separation of order of a parsec. The binary will undergo several other processes that affect its orbital decay, including stellar hardening and interactions with nuclear gas \citep[see][for a review]{Colpi2014}.
We modelled these additional delays with a single parameter, $t_{\rm delay}$, taken to be the same for all binaries. 
Our model could be extended by allowing the delay time to be drawn from a parametrised distribution that depends on the physical parameters of the system.

\subsubsection{Merger}

Once the total time delay had elapsed, binaries merged and formed remnants with mass given by the sum of the masses at the time of the merger. We neglected the mass loss due to the emission of GWs, but we did account for the accretion occurring from binary formation until merger, using the prescriptions described in the section below. We labelled the merger remnant as the primary BH of its host halo. 

We did not model kicks on the remnant BH due to the emission of GWs, which could displace the BH from the centre of its host, because they depend sensitively on the BH spins and the details of the orbit at the time of merger (eccentricity, spin inclinations), which we did not track in the current version of \texttt{POMPOCO}. Kicks are expected to have a small impact in high-mass galaxies, because of their large escape velocity. Those are the galaxies that contribute the most to the GW background in the nHz band and the high end of the LF, which are the observables of interest for this paper. Neglecting kicks is therefore expected to have little impact on our results. It might impact the LISA rates, however, because many LISA mergers occur in low-mass galaxies, and could modify the low mass end of the $M_*$-$M_{\rm BH}$ relation. We left the inclusion of kicks for future work.

{
\renewcommand{\arraystretch}{1.3}
\begin{table*}[hbtp!]
\caption{Parameters and settings of our model. }\label{tab:params}
\centering  
   \begin{tabular}{c *{3}{c|}}
   
   \cline{1-3}
   
  \multicolumn{1}{c}{Parameter}  &  \multicolumn{1}{|c|}{Description} &  \multicolumn{1}{|c}{Prior/Value} \\
   
    \cline{2-3}

    \hline
    
    \multicolumn{1}{c}{$M_{h,{\rm seed}} [M_{\odot}]$} & \multicolumn{1}{|c|}{minimum mass of the halos that host a seed BH} & \multicolumn{1}{|c}{$\log \mathcal{U}[10^6,10^{7.7}]$} \\

    \multicolumn{1}{c}{$f_{\rm seed}$} & \multicolumn{1}{|c|}{probability to seed a leaf halo with $M_h \geq M_{h,\rm {seed}}$ at $z\geq 10$} & \multicolumn{1}{|c}{$\log \mathcal{U}[0.01,1]$} \\

     \multicolumn{1}{c}{$\mu_{\rm seed} [ M_{\odot}]$  } & \multicolumn{1}{|c|}{mean of the log-normal distribution of seed masses} & \multicolumn{1}{|c}{$\log \mathcal{U}[10^{2.5},10^{6}]$} \\

     \multicolumn{1}{c}{$\sigma_{\rm seed} [{\rm dex}]$ } & \multicolumn{1}{|c|}{standard deviation of the log-normal distribution of seed masses} & \multicolumn{1}{|c}{$\log \mathcal{U}[10^{0.5},10^2]$} \\

      \multicolumn{1}{c}{$t_{\rm delay} [{\rm Myr}]$ } & \multicolumn{1}{|c|}{additional time delay (after DF timescale) for BH binary merger} & \multicolumn{1}{|c}{$\log \mathcal{U}[10^{-0.5},10^4]$} \\

      \multicolumn{1}{c}{$t_{\rm burst} [{\rm Myr}]$ } & \multicolumn{1}{|c|}{duration of burst accretion mode} & \multicolumn{1}{|c}{$\log \mathcal{U}[1,100]$}  \\

      \multicolumn{1}{c}{$\gamma_{\rm burst}$ } & \multicolumn{1}{|c|}{slope of the power-law distribution of $f_{\rm Edd,burst}$} & \multicolumn{1}{|c}{$\mathcal{U}[-1,0]$} \\

      \multicolumn{1}{c}{$\mu_{\rm steady} [{\rm dex}]$ } & \multicolumn{1}{|c|}{mean of the log-normal distribution of $f_{\rm Edd,steady}$} & \multicolumn{1}{|c}{$\mathcal{U}[-6,-3]$} \\

      \multicolumn{1}{c}{$\sigma_{\rm steady} [{\rm dex}]$ } & \multicolumn{1}{|c|}{standard deviation of the log-normal distribution of $f_{\rm Edd,steady}$} & \multicolumn{1}{|c}{$\mathcal{U}[0.5,3]$} \\

      \multicolumn{1}{c}{$z_{\rm cut}$ } & \multicolumn{1}{|c|}{redshift below which accretion is cut in the most massive BHs} & \multicolumn{1}{|c}{$\mathcal{U}[1,4]$} \\

      \multicolumn{1}{c}{$m_{\rm cut,0}[M_{\odot}]$  } & \multicolumn{1}{|c|}{mass at $z=0$ above which accretion is cut} & \multicolumn{1}{|c}{$\log \mathcal{U}[10^6,10^{7.3}]$} \\

      \multicolumn{1}{c}{$\alpha_{\rm cut}$ } & \multicolumn{1}{|c|}{slope of the increase in the mass cut for accretion with redshift, see Eq.~\eqref{eq:mass_cut_acc}} & \multicolumn{1}{|c}{$\mathcal{U}[0,0.5]$} \\
    
   \hhline{=|=|=}

    \multicolumn{1}{c}{$f_{\rm Edd,burst}$ } & \multicolumn{1}{|c|}{burst mode Eddington ratio} & \multicolumn{1}{|c}{$\mathcal{P L}$($\gamma_{\rm burst}-1$, $10^{-2}$, $10$)} \\

    \multicolumn{1}{c}{$f_{\rm Edd,steady}$ } & \multicolumn{1}{|c|}{steady mode Eddington ratio} & \multicolumn{1}{|c}{$\log \mathcal{N}(\mu_{\rm steady},\sigma_{\rm steady}) \leq 1$} \\

    \multicolumn{1}{c}{$t_{\rm steady}[{\rm Myr}]$ } & \multicolumn{1}{|c|}{duration of a steady accretion episode, after which $f_{\rm Edd,steady},t_{\rm steady}$ are redrawn } & \multicolumn{1}{|c}{$\log \mathcal{U}[1,100]$} \\

    \multicolumn{1}{c}{$\epsilon$ } & \multicolumn{1}{|c|}{radiative efficiency} & \multicolumn{1}{|c}{0.1} \\

     \multicolumn{1}{c}{$q_{h,\rm major}$ } & \multicolumn{1}{|c|}{mass ratio threshold for major halo mergers} & \multicolumn{1}{|c}{0.13} \\

     \multicolumn{1}{c}{$p_{\rm multi}$ } & \multicolumn{1}{|c|}{probability of interaction between three or more BHs} & \multicolumn{1}{|c}{0.22} \\

   \end{tabular}
   \tablefoot{The first 12 lines (before the double bar) list the free parameters of our model, which we fit to observations. The remaining lines summarise the distributions on quantities entering the simulations and the parameters that are fixed in our model. In the third column, ($\log$) $\mathcal{U}$ stands for a ($\log$-) uniform distribution in the given range, $\mathcal{P L}$ for a power-law distribution with given slope and range and $\log \mathcal{N}$ for a log-normal distribution with given mean and standard deviation.}
  \end{table*}

\subsection{Accretion}\label{sec:accretion}

We parametrised the growth of the mass of a BH, $m_{\rm BH}$, through accretion by the Eddington ratio, $f_{\rm Edd}$,
\begin{align}
    \dot{m}_{\rm acc}&=f_{\rm Edd} \dot{m}_{\rm Edd}, \\
    \dot{m}_{\rm BH}&=(1-\epsilon) \dot{m}_{\rm acc}, \\
    \dot{m}_{\rm Edd}&=\frac{m_{\rm BH}}{\epsilon t_{\rm Edd}} = \frac{L_{\rm Edd}}{\epsilon c^2},
\end{align}
where $\dot{m}_{\rm Edd}$ and $L_{\rm Edd}$ are the Eddington accretion rate and luminosity, respectively,
$\epsilon$ is the radiative efficiency, and $t_{\rm Edd}=450  \ {\rm Myr}$ defines the accretion timescale. 
In this way, the mass between two time steps increases as
\begin{equation}
    m_{\rm BH}(t+\Delta t)=m_{\rm BH}(t)\exp \left(f_{\rm Edd}\frac{1-\epsilon}{\epsilon}\frac{\Delta t}{t_{\rm Edd}} \right ).
\end{equation}
We took the radiative efficiency to be $0.1$. Inspired by \cite{Ricarte:2017ihq}, we considered two accretion modes: \emph{burst} and \emph{steady} accretion.

The burst mode was triggered following a major halo merger, $q_h \geq  q_{h,\rm major}$. Such mergers are expected to feed the reservoir surrounding MBHs, leading to an episode of intense accretion. When this happened, the most massive BH in the resulting galaxy started accreting for a time $t_{\rm burst}$ with an Eddington ratio $f_{\rm Edd, burst}$. In the burst mode, we did allow for super-Eddington accretion, and drew $f_{\rm Edd, burst}$ from a power-law distribution with slope $(\gamma_{\rm burst}-1)<0$ between $10^{-2}$ and 10. In this work, we considered a common burst time $t_{\rm burst}$ for all BHs. If the resulting halo underwent another major merger before a time $t_{\rm burst}$ had elapsed, we restarted the count. The choice of a power-law distribution with negative slope is motivated by the general expectation that super-Eddington accretion should be rare. \cite{Aird2018} found that X-ray observations of AGN activity across redshift and galaxy type suggest a steep power-law at high accretion rates, allowing for rare and short-lived periods of super-Eddington accretion.

Whenever they were not accreting in the burst mode, BHs accreted in the steady mode with an Eddington ratio $f_{\rm Edd, steady}$ drawn from a log-normal distribution with mean $\mu_{\rm steady}$ and variance $\sigma_{\rm steady}$ truncated at the high end at 1. When drawing $f_{\rm Edd, steady}$, we drew jointly $t_{\rm steady}$ from a log-uniform distribution between 1 and 100 Myr. After a time $t_{\rm steady}$ had elapsed, we drew again both quantities. This procedure is meant to capture the variability of AGNs. The choice of a log-normal distribution is motivated by \cite{Volonteri_2016}, Fig.~14, which shows the distribution of accretion rates in AGNs from the Horizon-AGN simulation~\citep{2014MNRAS.444.1453D}. Observations also suggest that accretion rates could follow a combination of log-normal and power law distributions, see \cite{Aird2018}. 

The most massive BHs are observed to be quiescent at low redshift due to gas having been consumed by star formation, as well as affected by supernova and AGN feedback \citep[see][and references therein]{2004MNRAS.353.1035M,Hirschmann:2012xp,Hirschmann:2013qfl}. To reproduce this, for $z\leq z_{\rm cut}$ we shut off accretion for BHs with $m_{\rm MBH} \geq m_{\rm cut}(z)$. We parametrised $m_{\rm cut}(z)$ as
\begin{equation}
    \log_{10}m_{\rm cut}(z)=\log_{10}m_{\rm cut,0}(1+z)^{\alpha_{\rm cut}}, \label{eq:mass_cut_acc}
\end{equation}
where $\log_{10}m_{\rm cut,0}$ and $\alpha_{\rm cut}$ are parameters of our model, together with $z_{\rm cut}$. We imposed $\alpha_{\rm cut}\geq 0$ so that the mass cut is larger at higher redshift. This approach gives our model the flexibility to capture the anti-hierarchical growth observed in data and simulations \citep{2004MNRAS.353.1035M,Hirschmann:2012xp,Hirschmann:2013qfl}, without enforcing it a priori\footnote{For instance, if the posterior of $z_{\rm cut}$ was to favour values close to 0, or the one of $m_{{\rm cut},0}$ values above the mass of most astrophysical BHs, for example $10^{11} M_{\odot}$, this would be a sign that the model does not need anti-hierarchical growth to fit the data.}.

We applied the prescriptions above to the primary and binary BHs of all haloes. Outer BHs do not accrete in our model, as we do not expect them to have a sufficient reservoir to accrete significantly. Both BHs in a binary had the same $f_{\rm Edd}$ (either burst or steady). 

}

 \section{Observables}\label{sec:observables}

In order to compare our model with observations we generated full MBH population for different choices of parameters. 
To do so, we applied the following procedure:
\begin{itemize}
    \item We built $N_h$ merger trees, drawing the mass of the haloes at $z=0$ from a log-normal distribution between $10^{11}$ and $10^{15} \ M_{\odot}$;
    \item We populated each merger tree according to the model parameters;
    \item We saved the properties of the BHs in the merger tree (mass, Eddington ratio, mass of the host halo, etc.) at a set of redshifts.
\end{itemize}
In this work, we generated $N_h=2000$ haloes. 

The generation of the DM halo merger tree is the most time-consuming step of our model. Fortunately, none of the parameters affect this step. We could therefore generate a single merger tree and run simulations for many sets of parameters on it. This resulted in a relatively fast model, and allowed us to thoroughly explore the parameter space. As an example, it takes $\sim 1$ hour to generate a merger tree and run the simulation for 500 sets of parameters on it. The $2000$ merger trees were run in parallel and the results were combined in post-processing. 

We will now describe how we transformed the output of our simulations into quantities that can be compared with observations: the LF, the stellar mass-black hole mass relation, the stochastic GW background, and the merger rate.

 \subsection{Luminosity function}

The luminosity of a BH is given by
 \begin{equation}
     L_{\rm BH}=f_{\rm Edd}\frac{m_{\rm BH}}{t_{\rm Edd}}c^2,
 \end{equation}
where $f_{\rm Edd}$ is either the burst mode or steady mode Eddington ratio, depending on the current accretion mode of the BH. We assigned binaries a luminosity equal to the sum of the luminosities of the individual BHs (recall that in our model, BHs in a binary share the same $f_{\rm Edd}$). Outer BHs of a halo were not included in the computation of the LF since we do not expect them to accrete significantly.

We denote the LF at redshift $z_0$ by $\Phi(L_{\rm BH},z_0)$. 
It is given by the weighted sum of the contribution from each halo merger tree, 
 \begin{equation}
  \Phi(L_{\rm BH},z_0)=\int W_{\rm PS}(M_{h,0})\frac{{\rm d}N_{M_{h,0}}}{{\rm d}\log_{10} L_{\rm BH}} \Big |_{z_0} \ {\rm d}M_{h,0},
 \end{equation}
where $W_{\rm PS}(M_{h,0})$ is the Press-Schechter weight for a halo of mass $M_{h,0}$ at $z=0$ and $\frac{{\rm d}N_{M_{h,0}}}{{\rm d}\log_{10}L_{\rm BH}}|_{z_0}$ is the number of BHs with luminosity $L_{\rm BH}$ at redshift $z_0$ in the merger history of a halo of mass $M_{h,0}$ at $z=0$. We computed the integral in a discrete way, by the PS-weighted sum of the individuals histograms of each tree. Binning in log-luminosity, we have for the LF in bin $k$
\begin{align}
     \Phi(L_{{\rm BH},k},z_0)=\frac{\mathcal{D}_{h}}{N_h \Delta_{L_{\rm BH}}} \sum_i W_{\rm PS}(M_{h,0,i})\frac{{\rm d}N_{i}}{{\rm d} \log_{10} L_{{\rm BH},k}}\Big |_{z_0} , \label{eq:phi_lum}
    \end{align}
with  $\Delta_{L_{\rm BH}}$ the size of the bin in $\log_{10} L_{\rm BH}$ and $\mathcal{D}_{h} = \log_{10}(M_{h,0,{\rm max}})-\log_{10}(M_{h,0,{\rm min}})$ coming from the normalisation of the log-uniform distribution used to draw the halo masses. The Poisson error in each bin is obtained by assuming that the histograms resulting from the different simulations are independent Poisson realisations. The variances thus sum up and the error is given by
\begin{align}
     \Delta \Phi(L_{{\rm BH},k},z_0)= \frac{\mathcal{D}_{h}}{N_h \Delta_{L_{\rm BH}}} \sqrt{\sum_i W_{\rm PS}(M_{h,0,i})^2\frac{{\rm d}N_{i}}{{\rm d}\log_{10}L_{{\rm BH},k}}\Big |_{z_0}}  \, . \label{eq:err_phi_lum}
    \end{align}

\subsection{Relation of the stellar mass to the black hole mass}\label{sec:mstar-mbh_comp}

In our code, we did not directly model the evolution of the stellar mass $M_{*}$ of a given halo. Instead, we used the halo mass-stellar mass relation of \cite{2010ApJ...710..903M} to convert $M_h$ into $M_*$, including the scatter in the relation, to account for the spread in the relation. We also accounted for the redshfit evolution of the relation when computing the $M_*-M_{\rm BH}$ relation at high $z$.  

For a given stellar mass at redshift $z_0$, we can compute the average black hole mass across haloes at that redshift as
\begin{align}
    M_{\rm BH}(M_{*},z_0)= \frac{\int W_{\rm PS}(M_{h,0})  \frac{{\rm d}N_{M_{h,0}}}{{\rm d}m_{\rm BH}{\rm d}M_*} \Big |_{z_0} m_{\rm BH}  \ {\rm d}m_{\rm BH} 
  \  {\rm d}M_{h,0}}
    {\int W_{\rm PS}(M_{h,0})  \frac{{\rm d}N_{M_{h,z}}}{{\rm d}M_*} \Big |_{z_0} {\rm d}M_{h,0}},
\end{align}
where $\frac{{\rm d}N_{M_{h,0}}}{{\rm d}m_{\rm BH}{\rm d}M_*} \Big |_{z_0}$ is the number of haloes at redshift $z_0$ that contain a BH with mass $m_{\rm BH}$ and whose stellar mass is $M_*$, in the merger history of a halo of mass $M_{h,0}$ at $z=0$.

In practice, we predefined a fixed binning scheme for the stellar masses, and then associated each halo hosting a BH to the corresponding $M_{*,k}$ bin based on the $M_h-M_*$ relation of \cite{2010ApJ...710..903M}. We have 
\begin{align}
    M_{\rm BH}(M_{*,k},z_0)= \frac{\sum_i W_{\rm PS}(M_{h,0,i}) \sum_j m_{\rm BH,i,j,k}}
    {\sum_i W_{\rm PS}(M_{h,0,i}) N_{i,k}},
\end{align}
where $m_{\rm BH,i,j,k}$ are the masses of the BHs in haloes at $z_0$ in the $i$-th simulation whose stellar mass is in the $M_{*,k}$ bin, and $N_{i,k}$ is the number of such haloes. In the case of binaries, we used the total mass. Outer BHs are not included in this computation.  

We computed the standard deviation of the $M_{\rm BH}-M_*$ relation as 
\begin{align}
    \Delta M_{\rm BH}(M_{*,k},z_0)= & \sqrt{ \frac{\sum_i W_{\rm PS}(M_{h,0,i}) \sum_j m_{\rm BH,i,j,k}^2}
    {\sum_i W_{\rm PS}(M_{h,0,i}) N_{i,k}}   - M_{\rm BH}(M_{*,k},z_0)^2}.
\end{align}
Technically, there should also be a Poisson error (on $N_i$), but we neglected it, as we take the bins in stellar mass to be big enough so that the dominant source of error is the variance of BH masses within that bin and not the counting.

\subsection{Stochastic gravitational-wave background}

We computed the stochastic GW signal in the PTA band of our simulations with the formalism of \cite{Phinney:2001di} and \cite{Sesana:2008mz}. In this work, we assumed binaries to be on quasi-circular orbits. We introduced the chirp mass of a binary with masses $m_1$ and $m_2$ as $\mathcal{M}_c=(m_1^3 m_2^3)/(m_1+m_2)^5$.  Denoting the characteristic GW amplitude at (observer) frequency $f$ as $h_c(f)$, we have
\begin{equation}
 h_c^2(f)=\frac{4G}{\pi f^2 c^2} \int  \frac{W_{\rm PS}(M_{h,0})}{1+z}\frac{{\rm d}N_{M_{h,0}}}{{\rm d}\mathcal{M}_c {\rm d}z} \frac{{\rm d}E_{\rm GW}(\mathcal{M}_c)}{{\rm d} \ln f_r} {\rm d}z {\rm d}\mathcal{M}_c {\rm d}M_{h,0},
\end{equation}
where $\frac{{\rm d}N_{M_{h,0}}}{{\rm d}\mathcal{M}_c {\rm d}z} $ is the number of haloes hosting a binary merger with chirp mass $\mathcal{M}_c$ at redshift $z$ in the merger history of a halo of mass $M_{h,0}$ at $z=0$, and $\frac{{\rm d}E_{\rm GW}(\mathcal{M}_c)}{{\rm d} \ln f_r}$ is the energy emitted in the source rest-frame at frequency $f_r=f(1+z)$. For quasi-circular binaries, the latter is given by
\begin{equation}
\frac{{\rm d}E_{\rm GW}(\mathcal{M}_c)}{{\rm d} \ln f_r} =\frac{(\pi \mathcal{M}_c G f_r)^{2/3} }{3}\mathcal{M}_c.
\end{equation}
The characteristic strain thus has a power-law frequency dependence 
\begin{equation}
    h_c^2(f)=\frac{4G^{5/3}}{3\pi^{1/3}c^2} f^{-4/3} \int  W_{\rm PS}(M_{h,0})\frac{{\rm d}N_{M_{h,0}}}{{\rm d}\mathcal{M}_c  {\rm d}z} \frac{\mathcal{M}_c^{5/3}}{(1+z)^{1/3}} {\rm d}z {\rm d}\mathcal{M}_c {\rm d}M_{h,0}. \label{eq:gw_bckgd}
\end{equation}
The amplitude is determined here by the distribution of mergers in a given simulation.
In practice, we performed a Monte Carlo integration, 
\begin{equation}
    h_c^2(f)=\frac{4G^{5/3}\mathcal{D}_{h}}{3\pi^{1/3}c^2} f^{-4/3} \sum_i W_{\rm PS}(M_{h,0,i}) \sum_j \frac{\mathcal{M}_{c,i,j}^{5/3}}{(1+z_{i,j})^{1/3}}, \label{eq:hc_pta}
\end{equation}
where the second sum is performed over all the mergers in the $i$-th merger tree. The Poisson error is then given by
\begin{equation}
    \Delta h_c^2(f)=\frac{4G^{5/3} \mathcal{D}_{h}}{3\pi^{1/3}c^2} f^{-4/3} \sqrt{\sum_i W_{\rm PS}(M_{h,0,i})^2 \sum_j \left ( \frac{\mathcal{M}_{c,i,j}^{5/3}}{(1+z_{i,j})^{1/3}} \right )^2}. \label{eq:err_hc_pta}
\end{equation}
Finally, following \cite{EPTA:2023xxk}, we define the root-mean-square residual as 
\begin{equation}
    \rho_c^2(f)=\frac{h_c^2(f)}{12 \pi^2 f^3} \Delta_f, \label{eq:rho2}
\end{equation}
where $\Delta_f$ is the frequency resolution (the inverse of the observation time). 

\subsection{Merger rate}\label{sec:rates}

We introduce $m_t=m_1+m_2$ as the total mass of an MBH binary and $q_b=m_1/m_2\geq 1$ its mass ratio. The rate of MBH mergers per unit $m_t$, $q_b$ and $z$ is computed as 
\begin{equation}
    \frac{{\rm d}N}{{\rm d }t {\rm d}m_t {\rm d}q_b {\rm d}z}= \int 4 \pi d_c(z)^2 c W_{\rm PS}(M_{h,0})\frac{{\rm d}N_{M_{h,0}}}{{\rm d}m_t{\rm d}q_b {\rm d}z}   \ {\rm d}M_{h,0}  , 
\end{equation}
where $d_c(z)$ is the comoving distance at redshift $z$ and $\frac{{\rm d}N_{M_{h,0}}}{{\rm d}m_t  {\rm d}q_b  {\rm d}z} $ is the number of MBH mergers with parameters $(m_t,q_b,z)$ in the merger history of a halo of mass $M_{h,0}$ at $z=0$, $t$ is the detector-frame time. Monte Carlo integration gives
\begin{equation}
    \frac{{\rm d}N}{{\rm d }t {\rm d}m_t  {\rm d}q_b  {\rm d}z}=\frac{\mathcal{D}_{h}}{N_h } \sum_{i}  4 \pi d_c(z)^2 c W_{\rm PS}(M_{h,0,i})\frac{{\rm d}N_{i}}{{\rm d}m_t  {\rm d}q_b  {\rm d}z}. 
\end{equation}
In practice, to generate samples of observed MBH merger events for an observation time $T_{\rm obs}$ we used the following procedure. We read the ($m_t,q_b,z$) values of all merger events in our trees, attributed them a Poisson rate
\begin{equation}
    \lambda(m_t,q_b,z)=\frac{4 \pi d_c(z)^2 c W_{\rm PS}(M_{h,0,i})\mathcal{D}_{h}}{N_h}, \label{eq:rate_event}
\end{equation}
and drew the number of such events from a Poisson distribution with expected number $\lambda(m_t,q_b,z)T_{\rm obs}$. 

The detector-frame rate as a function of a single parameter, for example $m_t$, was obtained by marginalising over the remaining parameters, and the total rate was obtained by marginalising over all parameters $(m_t,q_b,z)$ 
\begin{equation}
    \frac{{\rm d}N}{{\rm d }t}= \int 4 \pi d_c(z)^2 c W_{\rm PS}(M_{h,0})\frac{{\rm d}N_{M_{h,0}}}{{\rm d}m_t  \  {\rm d}q_b  \ {\rm d}z}   \ {\rm d}M_{h,0}  {\rm d}m_t {\rm d}q_b {\rm d}z, 
\end{equation}
see also Eq.~(2) of \citet{Hartwig_2019}.
Finally, Monte Carlo integration gives  
\begin{equation}
    \frac{{\rm d}N}{{\rm d }t} =\sum_{(m_t,q_b,z)}  \lambda(m_t,q_b,z). 
\end{equation}

\section{Fitting to observations}
\label{sec:fits}

As a first application of \texttt{POMPOCO}, we fitted jointly the LF across redshifts and the GW stochastic background.

\subsection{Observations}

\begin{figure*}
\includegraphics[width=0.99\linewidth]{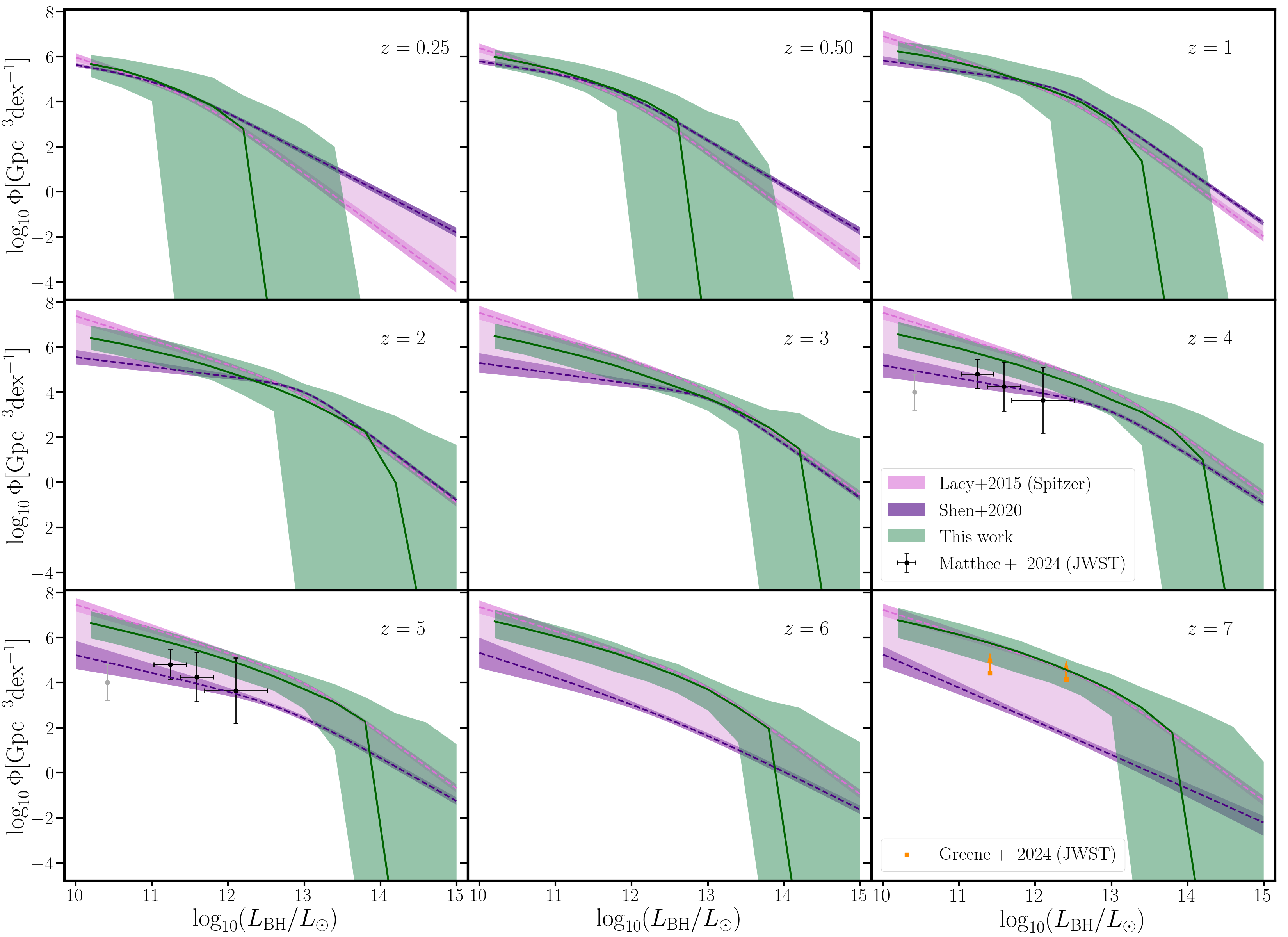}
   \caption{Evolution of the luminosity function. In green, we show the prediction of our model when fitting for the LF itself at redshifts $0.25$, 0.5, 3 and 6, as well as for the GW background measured by EPTA, displaying the median (green line) and 90\% confidence interval (green band). This should be compared with the fits to the observed LF from \cite{Lacy:2015tha} (pink) and \cite{Shen:2020obl} (purple) and the region in between (light pink), where we allow the observed LF to lie in our likelihood. At redshifts $z=4$, 5 we also plot constraints from JWST (squares) from \cite{Matthee2024}. The lowest luminosity point is shown in grey and not in black, because it is likely affected by incompleteness (see the discussion in \cite{Matthee2024}). At redshift $z=7$, we plot a lower bound obtained by \cite{Greene2024} using JWST results. 
   The recovered LF is remarkably consistent with observations also at redshifts at which we did not explicitly fit the data.
   }\label{fig:luminosity}
 \end{figure*}

As the observed LF, we considered a combination of the fits from \cite{Lacy:2015tha} (fit: `All') and \cite{Shen:2020obl} (fit: `global A'), as described in the next section (Fig.~\ref{fig:luminosity}).
The fit by \cite{Shen:2020obl} is based on a wide range of observations in the rest-frame mid-IR, B band, UV, soft and hard X-ray going up to $z\sim 7$. The constraint by \cite{Lacy:2015tha} is only based on the mid-IR Spitzer Space Telescope survey with observations up to $z \sim 3 $. It is also included in the compilation of \cite{Shen:2020obl}, although the latter is mainly driven by the more abundant B, UV, and X-ray data. The mid-IR LF fit in \cite{Lacy:2015tha}, if taken alone, suggests more AGN at a given bolometric luminosity. We further extrapolated the \cite{Lacy:2015tha} fit to higher redshifts and lower luminosities than covered by their data.

Although we did not use them to constrain our model, we show JWST results at $z=4, 5$ taken from \cite{Matthee2024} (as converted to bolometric luminosity by \cite{Habouzit2024}) in Fig.~\ref{fig:luminosity}. Data from \cite{Matthee2024} was broad-line selected from the FRESCO \citep{Oesch2023} and EIGER surveys, and span $z=4.2-5.5$. FRESCO and EIGER are spectroscopic slitless surveys with uniform selection in their fields, providing a complete sample with clear volume estimates. These results also hint towards a higher LF at high redshift than expected from the fit of \cite{Shen:2020obl}, in particular when taking into account that JWST is likely detecting only a subpopulation of all AGNs, and the obscured fraction could be as high as $80\%$ \citep{2022A&A...666A..17G}. We differentiate the lowest luminosity data point by plotting it in grey, as it is likely affected by incompleteness, see discussion in \cite{Matthee2024}.

Other JWST observations \citep{harikane2023jwstnirspeccensusbroadlineagns,Maiolino:2023bpi,2025A&A...697A.175S,2025ApJ...986..165T} suggest an even larger population of faint AGN, although sample selections are not as clean as in slitless surveys. In Fig.~\ref{fig:luminosity}, we show a lower bound to the LF at $z=7$ obtained by \cite{Greene2024} using continuum-based selections. These results point in the direction of an even higher LF, but it should be noted that this type of selection may include also non-AGN sources \citep{PerezGonzalez2024,Akins2024}.

For the stochastic GW background, we used the results of the second data release from EPTA \citep{EPTA:2023sfo,EPTA:2023akd,EPTA:2023fyk,EPTA:2023xxk,EPTA:2023gyr}, which are consistent with those of other PTAs \citep{InternationalPulsarTimingArray:2023mzf}, see Fig.~\ref{fig:pta}.      

In this work, we did not fit for the $M_*-M_{\rm BH}$ relation, but we perform a cross validation by comparing our results at $z=0$ with the fit from \cite{Greene2020} (`All', including scatter), which is based on a large compilation of galaxies. At higher redshifts, we compare to the fits from \cite{Pacucci:2023oci}, derived from JWST data at redshifts $z=$ 4 to 7.

\subsection{Computation of the likelihood and sampling}\label{sec:fitting_strategy}

Given the current uncertainty on the LF, we wished to explore a scenario where the true LF lies somewhere between the fits proposed in \cite{Shen:2020obl} and  \cite{Lacy:2015tha}.
We denote $\Phi_{\rm Lacy}(L_{\rm BH},z_0)$ and $\Phi_{\rm Shen}(L_{\rm BH},z_0)$ the mean of the fits from \cite{Lacy:2015tha} and \cite{Shen:2020obl} respectively, and $\sigma_{\rm Lacy}(L_{\rm BH},z_0)$ and $\sigma_{\rm Shen}(L_{\rm BH},z_0)$ their standard deviations (in log space). For a given redshift, we defined a probability distribution $p_{\rm LF}(\log_{10} \Phi,L_{\rm BH},z_0)$ that, in each luminosity bin, is flat between ${\rm min}(\log_{10}\Phi_{\rm Lacy}-\sigma_{\rm Lacy},\log_{10}\Phi_{\rm Shen}-\sigma_{\rm Shen})$ and ${\rm max}(\log_{10}\Phi_{\rm Lacy}+\sigma_{\rm Lacy},\log_{10}\Phi_{\rm Shen}+\sigma_{\rm Shen})$ and beyond these limits falls as a Gaussian with standard deviation $\sigma_{\rm Lacy}$ or $\sigma_{\rm Shen}$ accordingly. 

Our likelihood for the LF of a given set of parameters $\Lambda$, $\mathcal{L}_{\rm LF}(\Lambda,L_{\rm BH},z_0)$, was obtained by including the Poisson error on the LF computed from simulations so that
\begin{equation}
    \mathcal{L}_{\rm LF}(\Lambda,L_{\rm BH},z_0) =\int p(\Phi|\Lambda) p_{\rm LF}(\log_{10} \Phi,L_{\rm BH},z_0) \ {\rm d} \Phi,
\end{equation}
where $p(\Phi|\Lambda)$ is a normal distribution with mean given by Eq.~\eqref{eq:phi_lum} and standard deviation Eq.~\eqref{eq:err_phi_lum}. In practice, we performed a Monte Carlo integration, drawing samples for $\Phi$ from this normal distribution, truncating to positive values of $\Phi$.\footnote{Notice that we took the distribution on $\Phi$ to be normal, not on $\log_{10}\Phi$, since it is $\Phi$ that is a weighted sum of Poisson processes. Converting the error on $\Phi$ into an error on $\log_{10}\Phi$ through the standard propagation of errors is justified only when the error is much smaller than the quantity, which is not necessarily the case in the high luminosity bins where we have few counts, and would lead to underestimating the impact of Poisson errors. Ideally, we should use the formalism of \cite{Bohm:2013gla} rather than assuming a normal distribution when the total count is small and Poisson errors are large.} 
The total log-likelihood at $z_0$ was obtained by summing over the luminosity bins in a $z_0$-dependent range:
\begin{equation}
    \ln \mathcal{L}_{\rm LF}(\Lambda,z_0) =\sum _{ L_{{\rm BH,min}} \leq L_{\rm BH} \leq L_{{\rm BH,max}}} \ln \mathcal{L}_{\rm LF}(\Lambda,L_{\rm BH},z_0) \Delta_{L_{\rm BH}} \, .
\end{equation}
Finally, the LF log-likelihood was obtained by summing over the different redshifts we wished to fit: 
\begin{equation}
    \ln \mathcal{L}_{\rm LF}(\Lambda) =\sum _{z_0} \ln \mathcal{L}_{\rm LF}(\Lambda,z_0) \, .
\end{equation}

Similarly, we used the EPTA posteriors \citep{EPTA:2023sfo,EPTA:2023akd,EPTA:2023fyk,EPTA:2023xxk,EPTA:2023gyr} on $\log_{10} \rho_c$, denoted by $p_{\rm PTA}(\log_{10} \rho_c)$, to define the PTA likelihood in a given frequency bin $f_0$ as
\begin{equation}
    \mathcal{L}_{\rm PTA}(\Lambda,f_0) =\int p(\log_{10}(h_c)|\Lambda) p_{\rm PTA}(\log_{10} \rho_c) \ {\rm d} \log_{10} \rho \, . 
\end{equation}
We took $p(\log_{10}(h_c)|\Lambda)$ to be a normal distribution and the mean and error were computed from Eqs.~\eqref{eq:rho2}, \eqref{eq:hc_pta}, \eqref{eq:err_hc_pta}.\footnote{In this case, we assumed a normal distribution on $\log_{10}\rho_c$ because the errors on $\rho_c^2$ is small enough so that propagation of error into $\log_{10}\rho_c$ can be used.} We performed a Monte Carlo integration using the samples of $p_{\rm PTA}(\log_{10} \rho_c)$ taht were provided with the EPTA data release.

In this work, as a fiducial result, we fitted the LF at $z=0.25, \  0.5, \  3 \  {\rm and} \  6$. For $z_0=0.25$, we fited in the range of luminosities $[10^{10},10^{11.5}] L_{\odot} $, for $0.5$, in the range $[10^{10},10^{12}] L_{\odot}$, while for $z_0=3$ and $6$, we fited in the range $[10^{12},10^{14}] L_{\odot}$ and $[10^{11.5},10^{13.5}] L_{\odot}$ respectively. 

Since we assumed MBH binaries to be on circular orbits, the slope of the GW background is constrained to -2/3, see Eq.~\eqref{eq:gw_bckgd}. Fitting our model on the whole GW frequency range would therefore give the EPTA data a too large relative weight in the total fit without giving it the flexibility to better accommodate the GW spectrum as a whole. To facilitate comparison with previous works \citep{EPTA:2023xxk,Barausse:2023yrx} and because the way in which we estimate the PTA spectrum (assuming a normal distribution) is more accurate at low frequencies \citep{Lamb:2024gbh}, we focused on the frequency bin at $f_0=1/(10 \ {\rm yr})$. 

Finally, we assigned zero likelihood to simulations that produce MBHs at $z=0$ with masses exceeding $10^{12} M_{\odot}$, as these are unrealistically massive. In our model, such MBHs can form from heavy seeds, $\sim 10^6 M_{\odot}$, which later accrete in burst mode with Eddington ratios reaching up to 10. By $z\sim 7$, these MBHs typically have masses around $10^8$–$10^9 M_{\odot}$, and since we limit accretion only at lower redshifts, there is no mechanism to prevent them from continuing to grow, potentially reaching masses incompatible with observations. A more physically motivated accretion model -- based on the available mass reservoir, rather than on Eddington ratios and accretion times -- would likely prevent the formation of these excessively large MBHs. However, since these are extreme outliers in our simulations (i.e., they appear in very few of the simulations that survive fitting to the LF and/or GW spectrum and in small numbers), we adopted this simple approach for now, leaving the implementation of a more refined accretion description for future work.

The priors used on the parameters of \texttt{POMPOCO} are detailed in Table~\ref{tab:params}. We explored the parameter space with an MCMC using the \texttt{Eryn} sampler \citep{Karnesis:2023ras}. Since it is computationally cheap to run serially several sets of parameters for a given merger tree, we used many walkers ($\sim 500$) and ran the sampler only for a few steps (a few hundred). 

\section{Combined fit to the luminosity function and the gravitational wave background}\label{sec:results}

\begin{figure}
\centering
\includegraphics[width=0.99\linewidth]{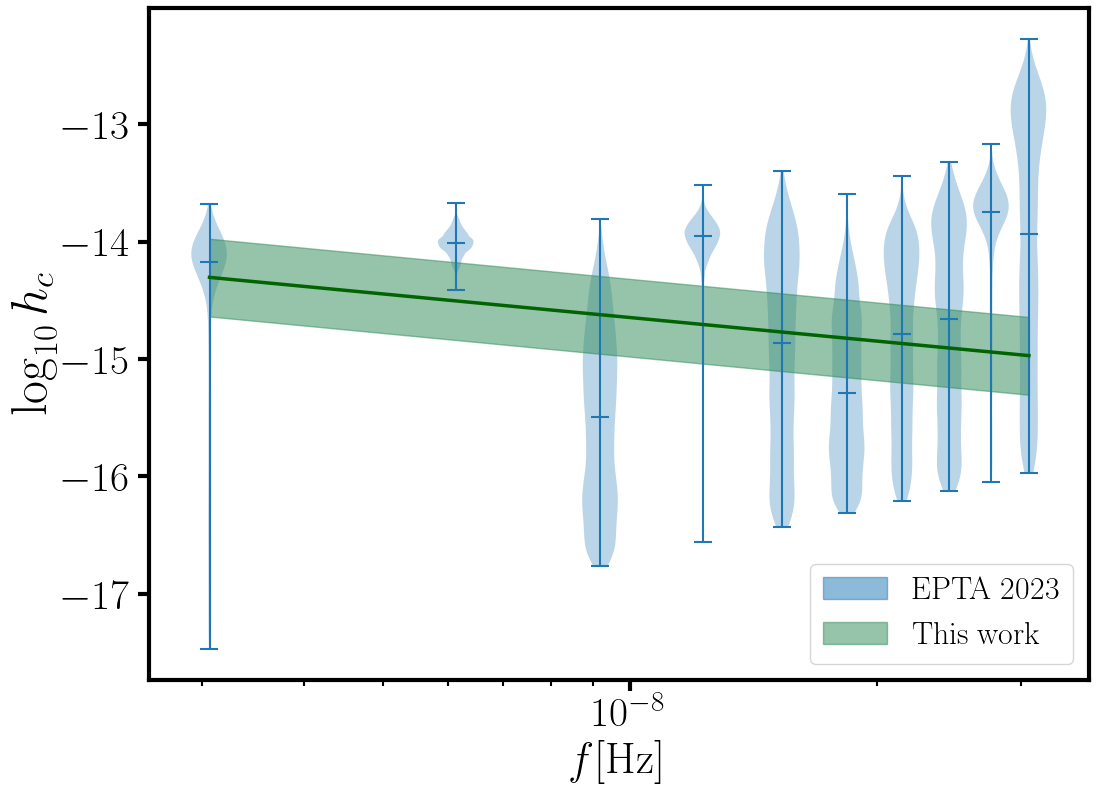}
   \caption{Stochastic GW background predicted by our model (green) and the free spectrum measured by EPTA (blue). The $90\%$ confidence intervals centred at the median for the amplitude of the GW background in the first and last bins i.e., at frequencies $1/(10 \ {\rm yr})$ and $1/(1 \ {\rm yr})$, are $4.8^{+5.1}_{-2.7}\times 10^{-15}$ and $1.0^{+1.1}_{-0.6}\times 10^{-15}$, respectively. We recall that we fit only for the first frequency bin and that we assume MBH binaries to be circular, so that the slope of the GW background spectrum is fixed to -2/3. Our result is compatible with the EPTA data, although it suggests a slightly lower median value. }\label{fig:pta}
 \end{figure}

 \begin{figure}
\centering
\includegraphics[width=0.99\linewidth]{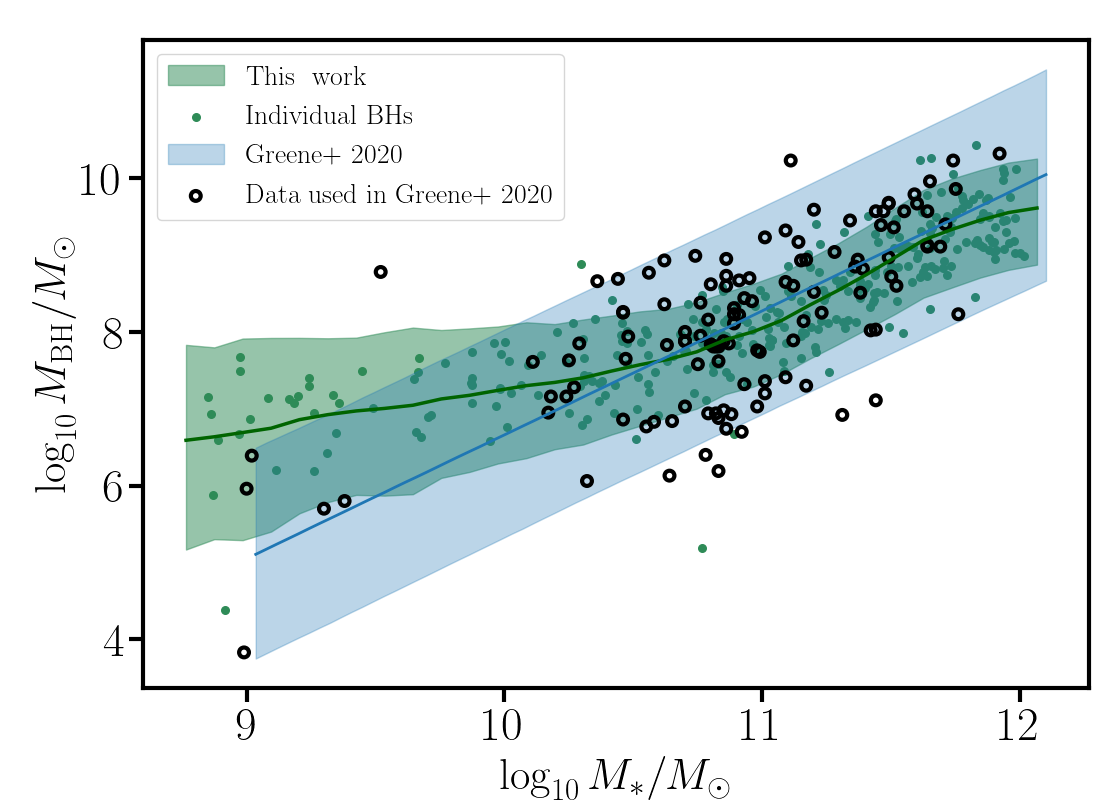}
   \caption{Relation of the stellar mass and MBH mass at redshift $z=0$. We compare our results, in green, with the fit of the local relation from \cite{Greene2020} (`all'), in blue, and the data also compiled in \cite{Greene2020}, in black. Green points show a (random) subsample of MBHs produced in our simulations, highlighting that our model can produce the most massive MBHs in haloes with $M_*\sim 10^{11} M_{\odot}$, while also producing MBHs of $10^7-10^{8.5} M_{\odot}$ in haloes with $M_* \geq 10^{11} M_{\odot}$. 
   }\label{fig:mstar_mbh}
 \end{figure}

Our results for the LF, the GW background and the $M_{*}$-$M_{\rm BH}$ relation at $z=0$ are shown in Fig.~\ref{fig:luminosity}, \ref{fig:pta} and \ref{fig:mstar_mbh} respectively. In each case, we present in green our model's median (solid line) and 90\% confidence interval (shaded region), as well as the observational data.   

In Fig.~\ref{fig:luminosity}, the light pink area denotes the region between the fits from \cite{Lacy:2015tha} (pink) and \cite{Shen:2020obl} (purple) where we allow the observational LF to reside. The recovered LF is remarkably consistent with observations at all redshifts -- including those for which we do not explicitly fit. Note that our LF vanishes at high luminosities because of the finite number of DM haloes we simulate, which does not allow us to generate the rare MBHs behind the very bright end of the LF. 

The GW spectrum (Fig.~\ref{fig:pta}) we recover aligns well with the EPTA data (in blue) though our median lies slightly lower. This suggests our results might be better compatible with the reduced amplitude of \cite{Goncharov:2024htb}, obtained by adopting an observationally driven model for the pulsar noise. To facilitate comparison to \cite{NANOGrav:2023hfp}, we report the median of the GW background at two reference frequencies, together with the errors defining a $90\%$ confidence region:
\begin{itemize}
    \item $h_c[1/(10 \ {\rm yr})]=4.8^{+5.1}_{-2.7}\times 10^{-15}$,
    \item $h_c[1/(1 \ {\rm yr})]=1.0^{+1.1}_{-0.6}\times 10^{-15}$.
\end{itemize}
As a comparison, if, instead of performing the fiducial fit to the LF and PTA described in Sec.~\ref{sec:fitting_strategy}, we use {\texttt{POMPOCO} to fit only for the LF of \cite{Shen:2020obl} across redshifts, we obtain the following estimates: 
\begin{itemize}
    \item $h_c[1/(10 \ {\rm yr})]=3.9^{+7.0}_{-2.5}\times 10^{-16}$,
    \item $h_c[1/(1 \ {\rm yr})]=8.0^{+14.3}_{-5.1}\times 10^{-17}$.
\end{itemize}
The median values in the latter case are more than one order of magnitude below the ones reported above. Within the model itself, there is not so much the possibility of having a much lower LF at high z and producing a high GW background. In physical terms, a higher LF means higher density of BHs and so this is why the model can match the EPTA results. 
We recall that we assume MBH binaries to be circular, therefore the slope of the GW background spectrum is fixed to be -2/3, see Eq.~\eqref{eq:gw_bckgd}. Including eccentricity and/or interactions with the environment would allow for a more flexible and shallower spectrum \citep{Enoki:2006kj,Bonetti:2017lnj,EPTA:2023xxk,Ellis:2023dgf}. We plan to explore these possibilities in future work.

The resulting $M_{*}$-$M_{\rm BH}$ relation at $z=0$ (Fig.~\ref{fig:mstar_mbh}) is in good agreement with the fit from \cite{Greene2020} (in blue), particularly in the region where most of the observational data (black points) are concentrated. A random sub-sample of individual MBHs from our simulations (green dots) shows that our model is capable of producing the most massive BHs observed in galaxies with $M_*\sim 10^{11} M_{\odot}$, while also producing MBHs of $10^7-10^{8.5} M_{\odot}$ in galaxies with $M_* \geq 10^{11} M_{\odot}$. While these MBHs are clear outliers, our model is able to reproduce them, a feature that some other models find challenging, as discussed in~\cite{2021MNRAS.503.1940H}.

\begin{figure*}[h]
\centering
\includegraphics[width=0.95\linewidth]{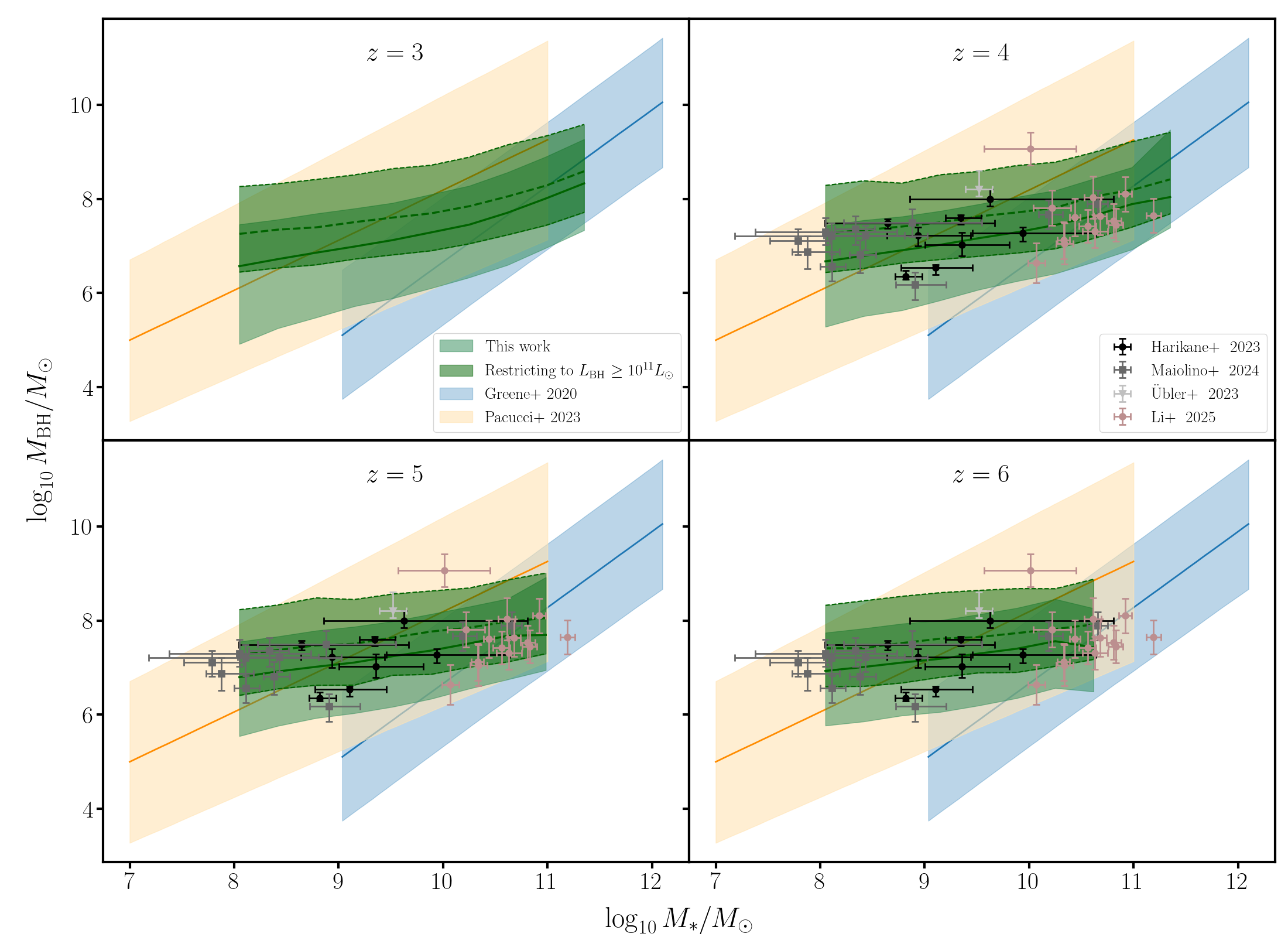}
   \caption{Relation of the MBH mass and stellar mass at redshift $z=$3, 4, 5 and 6. The darker green regions represent our results when restricting to MBHs with luminosity larger than $10^{11}L_{\odot}$, in order to mimic selection effects at high redshift. 
   For $z=$ 4, 5 and 6, we show JWST observations in the redshift range $z=$ 4 to 7 \citep{Maiolino:2023bpi, 2023A&A...677A.145U,harikane2023jwstnirspeccensusbroadlineagns,2025arXiv250205048L}, some of which were used to obtain the fit of~\cite{Pacucci:2023oci} (light orange, excluding the more recent data of~\cite{2025arXiv250205048L}). For comparison, we also show the fit to the observed local relation (light blue) of~\cite{Greene2020}.}\label{fig:mstar_mbh_highz}
 \end{figure*}

Our model predicts a break in the $M_*-M_{\rm BH}$ relation for galaxies with mass $M_*< 10^{10.5} M_{\odot}$. This originates from the break in the $M_h-M_{*}$ relation. The relation predicted directly by our model, $M_{h}-M_{\rm BH}$, does not show such a break at small MBH masses (see Fig.~\ref{fig:mhalo_mbh} in the Appendix). This break is rather introduced by our use of the $M_{h}-M_{*}$ relation of~\cite{2010ApJ...710..903M}, which shows one precisely at $M_h\sim10^{12}M_{\odot}$, corresponding to $M_*~\sim 10^{10.5}M_{\odot}$. Such a break is also observed in some cosmological simulations~\citep{2021MNRAS.503.1940H}, and is imputed to the transition between SN feedback, suppressing MBH accretion in low-mass galaxies and AGN feedback, which leads to self-regulation of MBH accretion in high-mass galaxies and quenches star formation. For $M_*< 10^{10} M_{\odot}$, the median MBH mass is of order $10^{7} M_{\odot}$ and lies above observations, with a shallower slope. We note, however, that there are only a few observations with $M_{*}<10^{10}M_{\odot}$, so that the fit of \cite{Greene2020} is dominated by massive galaxies, and might not be accurate for low-mass galaxies. Ultimately, most observations in low-mass galaxies are within our $90\%$ confidence region, and even those outside can actually be reproduced as outliers of our distribution (green points). In any case, we have checked that this potential overestimate of MBHs with mass $10^{7} M_{\odot}$ does not impact our results on the PTA background, which is fully dominated by MBHs with mass $>10^{8} M_{\odot}$.

At the massive galaxies end, our $M_{*}-M_{\rm BH}$ relation shows a break at $M_* \gtrsim10^{12}M_{\odot}$ and slightly underestimates the mass of MBHs with respect to the observational fit. We note, however, that there is only one data point in this region ($M_* \gtrsim 10^{12}M_{\odot}$), so the fit might not be valid for these galaxies. As discussed in App.~\ref{app:mhalo_mbh}, we have verified that this potential underestimation of the MBH masses should have a small impact on the stochastic GW background. In short, the reason is that this affects MBHs in more massive haloes, which are heavily suppressed by the Press-Schechter mass function, and therefore contribute little to the background.

We show the $M_{*}-M_{\rm BH}$ relation at higher redshift in Fig.~\ref{fig:mstar_mbh_highz}. For $z=$ 4, 5 and 6, we plot the AGN candidates from JWST observations in the redshift range $z=$ 4 to 7 reported by~\cite{harikane2023jwstnirspeccensusbroadlineagns,2023A&A...677A.145U,Maiolino:2023bpi,2025arXiv250205048L}. Since the $M_{\rm h}$–$M_*$ relation from~\cite{2010ApJ...710..903M} is calibrated for $M_* \gtrsim 10^8 M_{\odot}$, we restrict our plot to this stellar mass range. Additionally, the redshift evolution of their fit is only calibrated up to $z\sim3.5$, so our results at higher redshifts are based on extrapolation. In order to mimic selection effects in this dataset, we also show in dark green our relation when restricting to MBHs with $L_{\rm BH}\geq 10^{11}L_{\odot}$.
Most data points are within the range predicted by our model when imposing this cut. Note that, when computing the $M_{*}-M_{\rm BH}$ relation predicted by \texttt{POMPOCO}, we also count MBHs in a binary as a single BH using the total mass of the binary, as described in Sec.~\ref{sec:mstar-mbh_comp}, because it is usually challenging for observations to distinguish a single MBH from a binary at close separations, although some candidates have beed identified \citep{Maiolino:2023bpi}. 
We also show the fit to the local relation of~\cite{Greene2020}, and the fit of~\cite{Pacucci:2023oci}, which is based on the same high redshift data points shown in Fig.~\ref{fig:mstar_mbh_highz} (except for the latest from~\cite{2025arXiv250205048L}). 
We caution the reader that the underlying intrinsic $M_{*}-M_{\rm BH}$ relation at high redshift, once selection biases are
accounted for, is still debated \citep{Pacucci:2023oci,Li2024Tip}, as we do not know yet if a large population of undetected MBHs with lower mass ratio between $M_{\rm BH}$ and $M_{*}$ exists. 

Overall, our model predicts a population in the same locus as the observations, although the slope is shallower than the fits of \cite{Greene2020} and  \cite{Pacucci:2023oci}. Indeed, our model predicts a sizeable population of MBHs with mass almost comparable with the host galaxy mass in low-mass galaxies. We leave a deeper investigation of the possible causes for this discrepancy to future work. Conversely, the model does not include many very high-mass BHs ($>10^9$ \Msun): this is related again to our simulated merger trees, which do not include the massive and rare haloes hosting 
the MBHs powering quasars at very bright end of the LF. 
At $z=$3 and 4, we can see that our model starts ``building up" to the steeper local $M_{*}-M_{\rm BH}$ relation of~\cite{Greene2020}.

  \begin{figure*}
\centering
\includegraphics[width=0.85\linewidth]{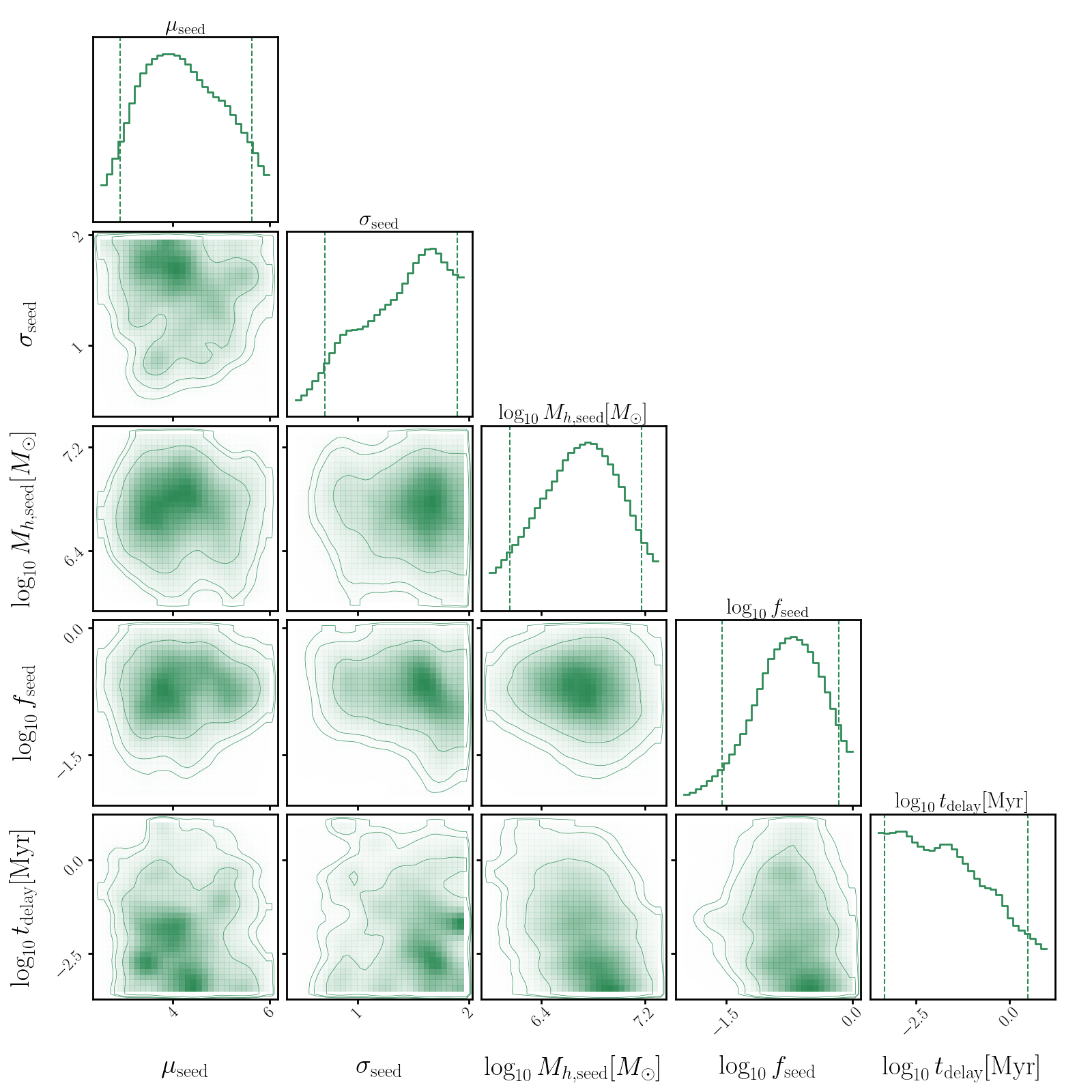}
   \caption{Posterior on the model parameters related to seeding and BH mergers, when fitting for the LF and GW background. We show: the mean and standard deviation of the log-normal distribution of seed BH masses, $\mu_{\rm seed}$ and $\sigma_{\rm seed}$; the minimum mass of haloes seeded $M_{h,\rm seed}$ and the seeding probability $f_{\rm seed}$; the delay of binary BH mergers (in addition to halo dynamical friction), $t_{\rm delay}$.}\label{fig:posterior_hyper_seed_delay}
 \end{figure*}

  \begin{figure*}
\centering
\includegraphics[width=0.95\linewidth]{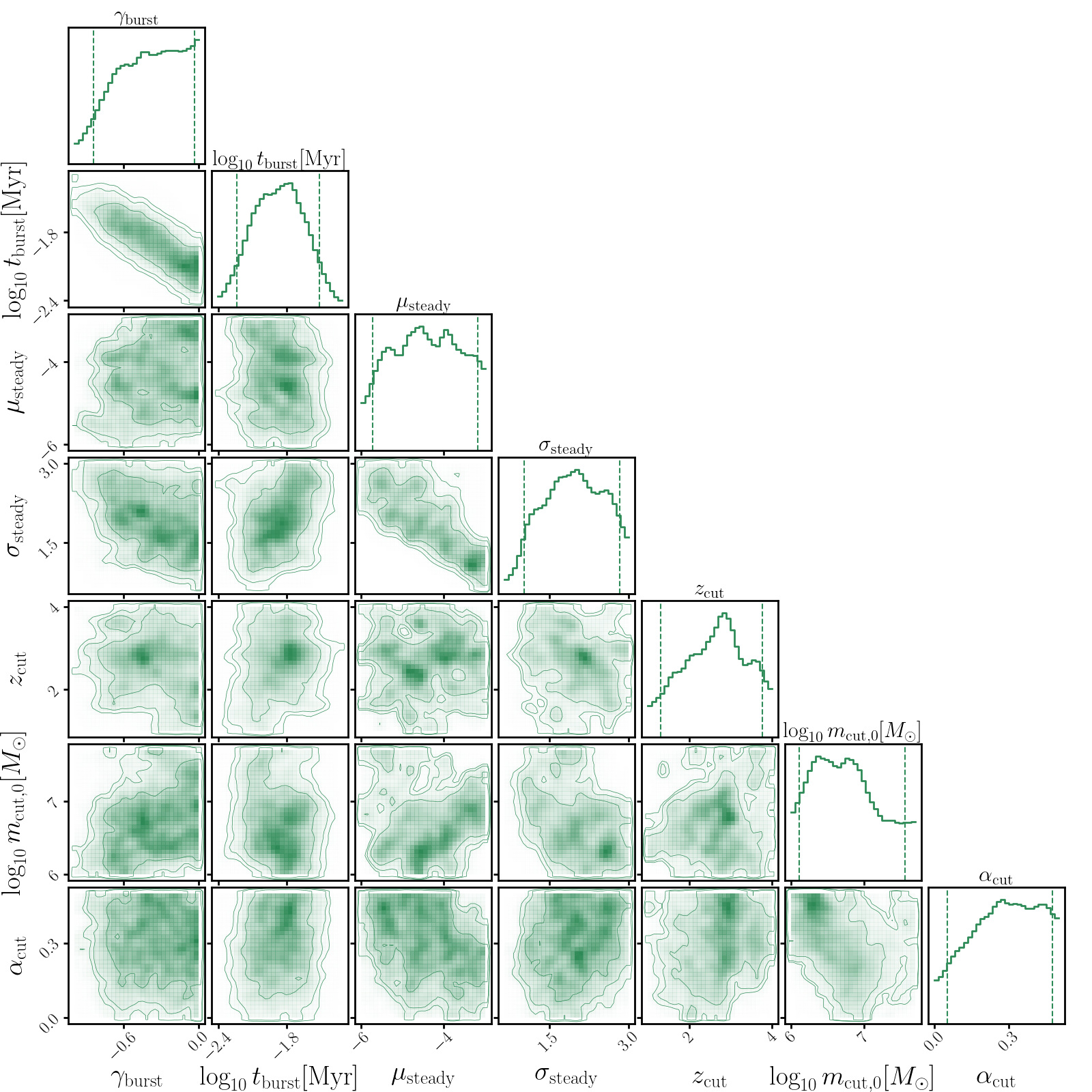}
   \caption{Posterior on the model parameters related to accretion, when fitting for the LF and the GW background. We show: the slope of the power-law distribution $\gamma_{\rm burst}$ and the duration $t_{\rm burst}$ of burst accretion; the mean and standard deviation of the log-normal distribution of steady accretion rates, $\mu_{\rm steady}$ and $\sigma_{\rm steady}$;
   the redshift $z_{\rm cut}$ below which accretion is shut off for heavier MBHs; the slope of the increase of the shut-off mass, $m_{\rm cut,0}$ at $z=0$, with redshift, $\alpha_{\rm cut}$.} \label{fig:posterior_hyper_accretion}
 \end{figure*}
 
 Overall, our analysis demonstrates that observations of the LF and of the (likely) GW background observed by PTAs can find a consistent description in our physical model, provided we allow for a more abundant population of AGN compared to global fits of the AGN LF based mainly on optical/UV/X-ray, pre-JWST data.

We show the posterior distribution of the parameters of our model in Fig.~\ref{fig:posterior_hyper_seed_delay} and~\ref{fig:posterior_hyper_accretion}. For the sake of readability, the first figure shows the parameters controlling the seeding and merger delays, and the second figure those controlling accretion. We do not observe strong correlations between these two groups of parameters. 

We obtain a relatively good constraint on the properties of haloes that are seeded, coming from the abundance of BHs in the Universe at different redshifts (i.e., the normalisation of the LF). 
Our model suggests that $\sim 10\%$ of haloes with $M_h \gtrsim 10^7 M_{\odot}$ at $z \sim 20$ should host an MBH seed.\footnote{These values should be interpreted with caution, in particular for $f_{\rm seed}$, due to the finite resolution of our DM halo merger trees. The minimum mass we resolve at $z=20$ is $\sim 2\times 10^6 (M_{h,0}/10^{14}) M_{\odot}.$ For the most massive halos with a mass of $10^{15} M_{\odot}$ at $z=0$, the minimum resolved mass therefore is above our best estimate for $M_{h,{\rm seed}}$. Because we allowed all leaves above $z=10$ to be seeded, however, haloes that fell below the mass resolution before $z=20$ might still be seeded. Moreover, such heavy haloes are strongly suppressed by the Press-Schechter mass function and contribute little to our Universe. Therefore, we expect our estimates to be qualitatively meaningful, in particular for $M_{h,{\rm seed}}$.} The range of values favoured for $M_{h,{\rm seed}}$ is more compatible with the prediction of relatively heavy seed models for the formation of seeds \citep[see][for an overview and discussion]{Regan2024}.
We obtain a broad posterior on the mean seed mass $\mu_{\rm seed}$, but with a clear preference for $\mu_{\rm seed}\gtrsim10^3 M_{\odot}$, in better agreement with the predictions of non-extreme heavy seed models \citep{Volonteri2021review}. The tendency towards large values of $\sigma_{\rm seed}$ suggests that we tend to prefer broad initial distributions for the mass seed. 
This is likely due to the need to accommodate observations spanning a broad range of MBH masses, together with the restrictive shape of the seed mass distribution currently adopted in \texttt{POMPOCO} (a single Gaussian). Allowing for a combination of seeding channels (e.g. each having its own $\mu_{\rm seed}$ and $\sigma_{\rm seed}$) might improve this, but at the cost of increasing the parameter space. We expect that LISA will allow us to better identify the contribution of different seeding channels compared to current observations \citep{Gair:2010bx,Sesana:2010wy,Klein:2015hvg,Toubiana:2021iuw}. 

We find large delays in MBH mergers to be strongly disfavoured, with $t_{\rm delay} \lesssim 1 {\rm Gyr}$ at $90\%$ confidence. We recall this is an additional delay to the DF timescale following halo mergers and that we have considered a single value for all MBHs, without a mass or redshift dependence. Our findings are in line with those of \cite{EPTA:2023xxk,Barausse:2023yrx,NANOGrav:2023hfp}, which found that PTA observations favour MBH binaries merging efficiently following galaxy mergers.

Moving on to the accretion parameters, the posterior on $\gamma_{\rm burst}$ suggests a log-flat distribution for $f_{\rm Edd}$ in burst mode following major halo mergers, favouring the occurrence of super-Eddington accretion episodes. These burst episodes are estimated to last a few tens of Myr, in agreement with the findings from \cite{2006ApJS..163....1H,2015MNRAS.447.2123C}. On average, burst accretion episodes increase the MBH mass by a factor of 1.2, rising to 2.8 in the case of super-Eddington episodes.
The parameters of steady-mode accretion, in turn, are poorly constrained individually. The anti-correlation between $\mu_{\rm steady}$ and $\sigma_{\rm steady}$ means, however, that $\sim 10\%$ of the MBHs that accrete in this mode have $f_{\rm Edd}> 10^{-2}$ and $ \sim 4\%$ have $f_{\rm Edd}> 10^{-1}$. When we label these as AGN, this suggests that the AGN fraction is in the range $4-10\%$ at ``cosmic noon'', which in our model corresponds to $z\sim z_{\rm cut}$. 
This is in good agreement with observations in different parts of the electromagnetic spectrum, which suggest an AGN fraction at $z\sim 1-3$ in the range $1-10\%$ \citep{2012ApJ...746...90A,2013ApJ...763..133T,2012ApJ...748..142D}. The strong correlations between $\gamma_{\rm burst}$ and $t_{\rm burst}$ and between $\mu_{\rm steady}$ and $\sigma_{\rm steady}$ point to the fact that a more physically driven approach to describe accretion, for instance based on a reservoir tracking the amount of mass available to accrete, might be possible, allowing to reduce the number of parameters in our model. We leave this exploration for future work. 

Finally, our results do support the anti-hierarchical hypothesis for the growth of MBHs, as  we observe a clear preference for accretion to shut off below $z_{\rm cut}\sim 2.5$ with $m_{{\rm cut},0}\sim 6\times 10^{6}M_{\odot}$. Although not very well constrained, the evolution parameter for the cut-off mass with redshift, $\alpha_{\rm cut}$, does seem to favour strictly positive values, leading to the estimate that accretion is suppressed at $z_{\rm cut}\sim 2.5$ for MBHs with mass $\gtrsim 10^8M_{\odot}$. The inferred value of $z_{\rm cut}$ is in good agreement with observations and more complete numerical simulations, which estimate the peak of accretion to be in the range $z=2-3$ \citep{2004MNRAS.353.1035M,Hirschmann:2012xp,Hirschmann:2013qfl}. Using our $M_*-M_{\rm BH}$ relation, we see that the preferred value of $m_{{\rm cut},0}$ corresponds to a stellar mass of $\sim 8 \times 10^{10} M_{\odot}$, in good agreement with where the break in the $M_h-M_*$ relation of \cite{2010ApJ...710..903M} lies, for $M_h\sim 10^{12} M_{\odot}$. This break has been interpreted as being due to the combined effect of the feedback from supernovae (in smaller haloes) and AGNs (in larger haloes) \citep{2006MNRAS.365...11C,Shankar:2006xz,2006MNRAS.370..645B,2008MNRAS.391..481S}. 

The fact that the inferred values of the parameters entering \texttt{POMPOCO} are overall in good agreement with findings from observations and/or more complete numerical simulations reinforces our confidence that our model is able to capture the key features driving the formation and evolution of MBHs and that our conclusions on the compatibility of the LF and the GW background are robust.

 \begin{figure}
\centering
\includegraphics[trim={.4cm 0cm 0 0},clip, width=0.98\linewidth]{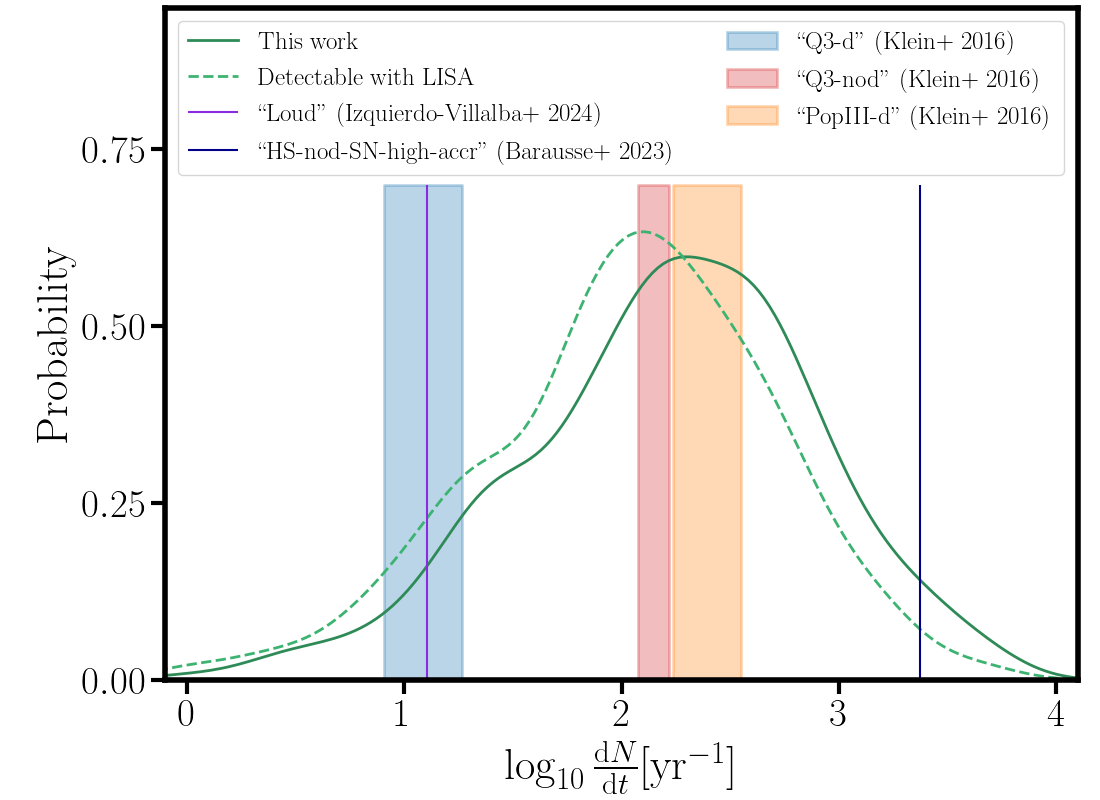}
   \caption{Prediction of the yearly rate of MBH mergers, after fitting for the LF and the GW background measured by EPTA. The solid curve shows the intrinsic rate, the dotted one the rate of detetable events, assuming an SNR threshold of 10. For comparison, we also report the predictions for the intrinsic rate of other models from the literature (see text for description). For ``Q3-nod'', ``Q3-d'' and ``PopIII-d'', the shaded areas show the range reported in Ref.~\cite{Barausse:2023yrx} (the lower bound
   corresponding to finite resolution results, and the upper bound to results extrapolated to infinite resolution). For model ``HS-nod-SN-high-accr'',
   we only report finite resolution results (i.e. a lower bound) as the extrapolation was not provided in Ref.~\cite{Barausse:2023yrx}.}\label{fig:lisa_rate}
 \end{figure}

  \begin{figure}
\centering
\includegraphics[trim={.cm 0cm 0 0},clip, width=0.98\linewidth]{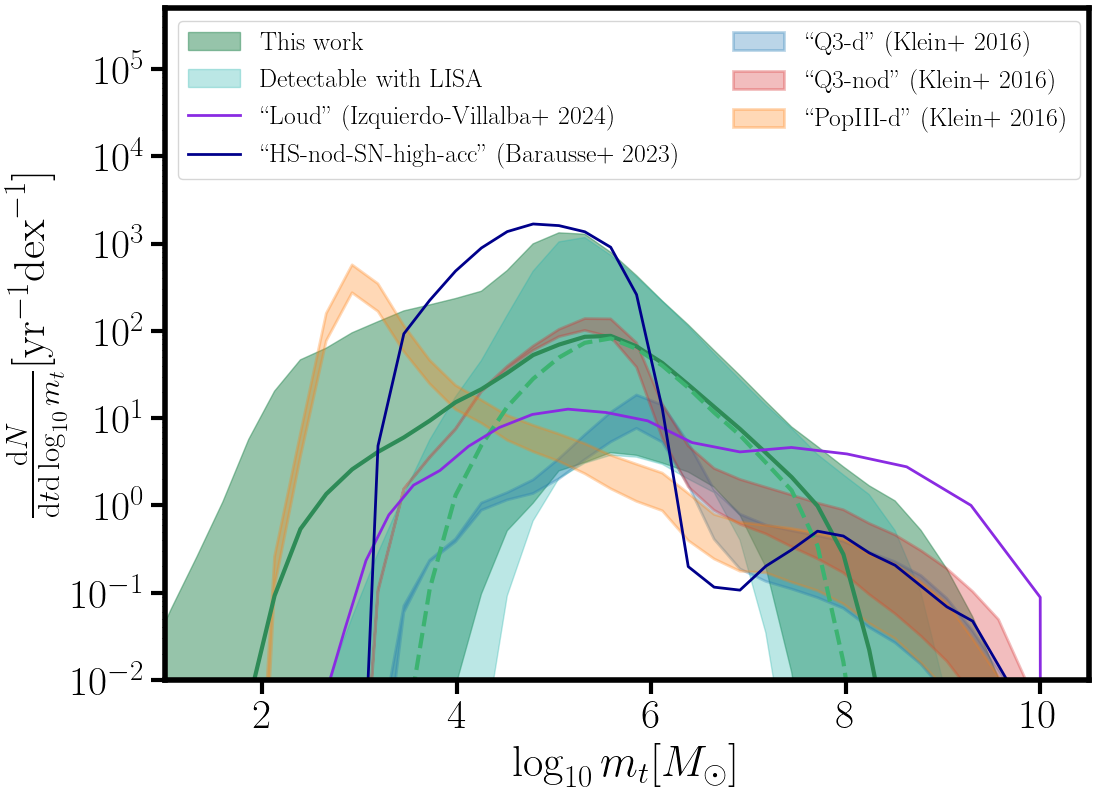}
   \caption{Prediction of the yearly rate of mergers as a function of the total source-frame mass of the binary. Thick lines show the median and shaded areas the $90\%$ confidence region after fitting for the LF and the GW background of EPTA. The green curves correspond to the intrinsic merger rate, while the light green curves show the rate of events detectable by LISA, assuming an SNR threshold of 10. Predictions from other models are included for comparison.  }\label{fig:lisa_rate_mass}
 \end{figure}

  \begin{figure}
\centering
\includegraphics[trim={.4cm 0cm 0 0},clip, width=0.98\linewidth]{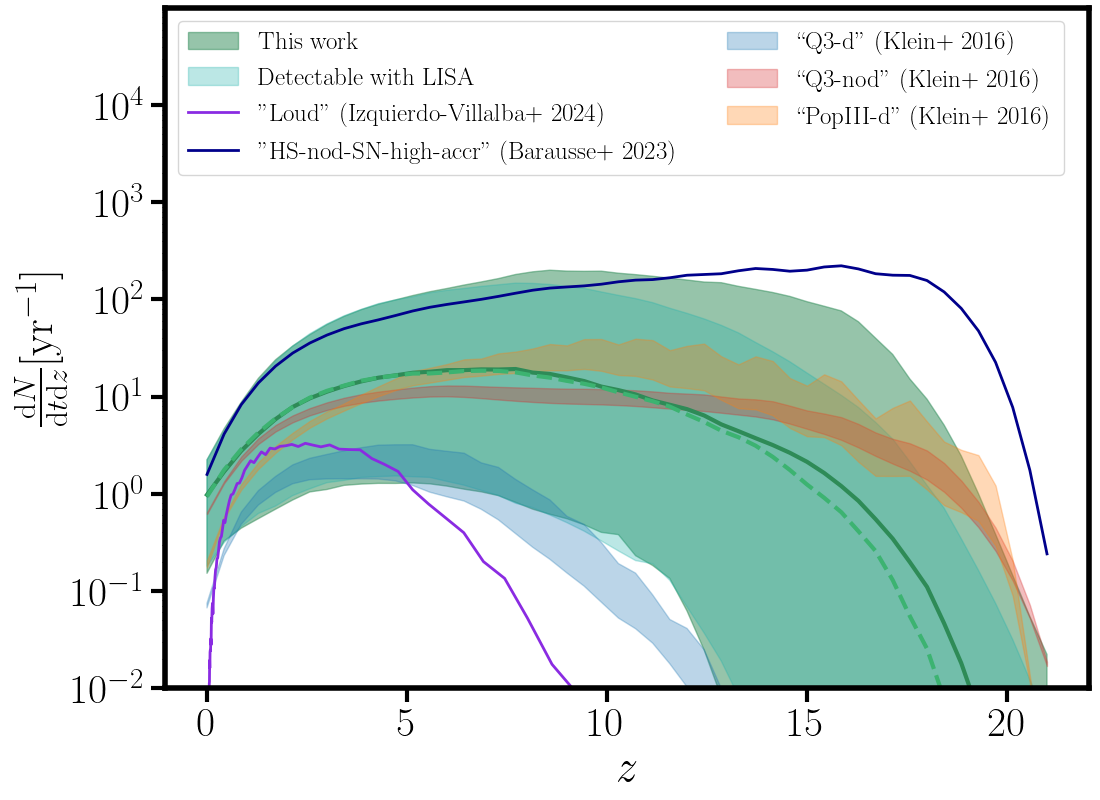}
   \caption{Prediction of the yearly merger rate as a function of the redshift (see also Fig.~\ref{fig:lisa_rate_mass}).}\label{fig:lisa_rate_z}
 \end{figure}

\section{Merger rates and LISA predictions}\label{sec:lisa}

Next, we turn to the predictions for merger rates and LISA observations based on our fits to the LF and the GW stochastic background. We compute the yearly rate as a function of MBH binary parameters, as well as the total rate (per year), following the procedure described in Sec.~\ref{sec:rates}. To estimate the LISA detection rate, we restrict to events with signal-to-noise ratio (SNR) above 10, which we assume to be the threshold for detection in a 4-year mission. The SNR of the sources is computed assuming the "Proposal" noise curve \citep{amaroseoane2017laserinterferometerspaceantenna} and using the IMRPhenomXHM model \citep{Garcia-Quiros:2020qpx} for the waveform. Since we do not track the evolution of spins in our simulations, for simplicity, we attribute a spin 0 to both BHs in a binary when computing SNRs. We do not expect our results for the expected number of detections to be much affected by this choice, since spins, at the level of the whole population, introduce a second order correction to the detectability of the sources. For each event, we draw the sky location, the phase at coalescence, the polarisation and the inclination angle isotropically.

Figure~\ref{fig:lisa_rate} shows the posterior distribution of the merger rate per year, while Figs.~\ref{fig:lisa_rate_mass} and~\ref{fig:lisa_rate_z} present this rate as a function of the total binary mass, $m_t$, and redshift, $z$, respectively. The thick lines represent the median values, with the shaded regions indicating the $90\%$ confidence intervals, and the thin lines illustrating individual realisations of our model. Green curves show the intrinsic rate, while the light green ones show the rate of events detectable with LISA. Our predictions for LISA cover a wide range, with the expected number of merger events varying from a few to thousands per year. 

In Figs.~\ref{fig:lisa_rate}, \ref{fig:lisa_rate_mass} and \ref{fig:lisa_rate_z} we also compare our results for the intrinsic rate with those of other models from the literature, selecting ones that have shown good agreement with EPTA \citep{EPTA:2023xxk}: models ``Q3-nod'', ``Q3-d'' and ``PopIII-d'' from \cite{Klein:2015hvg}, model ``HS-nod-SN-high-accr'' from \cite{Barausse:2023yrx}, and model ``Loud'' from \cite{2024A&A...686A.183I}. We refer to the upcoming paper by members of the LISA astrophysics working group for a more exhaustive comparison of LISA predictions in different SAMs and cosmological simulations~\citep{cat_paper}.

The models ``Q3-nod'', ``Q3-d'' and ``HS-nod-SN-high-accr'' assume that MBHs form from heavy seeds (from the collapse of protogalactic discs~\citep{Volonteri:2007ax}), while ``PopIII-d'' assumes that they form from the remnant of Pop III stars \citep{Madau:2001sc}. Models ``HS-nod-SN-high-accr'' and ``Q3-nod'' also feature no delays between
galaxy mergers and MBH mergers (although they include delays between halo and galaxy mergers), while ``PopIII-d'' and ``Q3-d'' do include delays due to stellar hardening, gas-driven migration and triple black-hole interactions that affect MBH binaries.
Moreover, in model ``HS-nod-SN-high-accr'' 
the influx of gas
into the nuclear regions hosting MBHs is boosted by a factor $\sim 4$, to achieve better agreement with the PTA results~\citep{EPTA:2023xxk,Barausse:2023yrx}.

The ``Loud'' model by \cite{2024A&A...686A.183I} uses the {\texttt{L-Galaxies} SAM with a mix of heavy and light seeds to track MBH binary evolution after galaxy mergers. It considers three phases: dynamical friction, hardening, and GW emission. The dynamical friction phase of the satellite MBHs lasts for a time given by~\cite{2008gady.book.....B} and starts at the radius at which the tidal forces stripped 80\% of the galaxy stellar mass. At the end of this phase, the MBHs form a gravitational bound binary, entering the hardening and GW inspiral phases whose separation and eccentricity are evolved consistently based on whether the environment is gas-rich or gas-poor. The model also accounts for triple interactions and, during the MBH binary growth, an anticorrelation between the accretion rate of the primary and secondary MBH (preferential accretion). Specifically, in the ``Loud'' model, the efficiency of MBH accretion is boosted to reproduce the GW background measured by EPTA.}

Our predictions agree best with the "Q3-nod" model, which assumes that MBHs form from heavy seeds and merge soon after galaxy mergers, which aligns well with the preferred values of $\mu_{\rm seeds}$ and $t_{\rm delay}$ in our analysis. 
The ``HS-nod-SN-high-accr'' model, which makes similar hypotheses for the seeding and the delay, also shows good agreement with our results, the main difference being its large merger rate at high redshift.\footnote{The larger merger rate at high redshift in model ``HS-nod-SN-high-accr'' is due to both the different threshold Toomre parameter used for seed formation in unstable protogalactic discs, and to better tracking of sub-resolution haloes hosting an MBH seed, see \cite{Barausse:2023yrx} for more details.}
As values of $\mu_{\rm seed}\sim 10^2 M_{\odot}$ are also allowed by our analyis, the ``PopIII-d'' model is also mostly compatible with our results, although it is not preferred. This confirms that LISA will play a crucial role in disentangling formation scenarios. 

The ``Loud" model predicts mergers only from $z \lesssim 10$, in tension with our results. This discrepancy may stem from the assumption in \texttt{POMPOCO} that the delay between galaxy and MBH mergers is independent of galaxy mass. In our model, this additional delay beyond the DF timescale is parametrised by $t_{\rm delay}$ is applied uniformly to all mergers. MBH binaries resulting from low-mass galaxy mergers are expected to experience longer formation and coalescence timescales than those in more massive systems, however. Supporting this hypothesis, the ``Q3-d'' model, which includes a post-galaxy-merger delay, yields a redshift distribution similar to the ``Loud'' model, whereas the ``Q3-nod'' model, which omits this delay, aligns more closely with our predictions. We plan to refine this aspect in future work. Note that this would not necessarily lead to a reduction in the PTA GW background, since delays following mergers of more massive galaxies—those hosting the MBHs that dominate the background could be relatively short. In this sense, our $t_{\rm delay}$ parameter may be more reflective of the delay timescales in such massive systems.
In terms of total mass, the predictions of the ``Loud'' model are generally consistent with our results across most of the mass spectrum, with the exception of the high-mass end. There, it predicts significantly more mergers involving MBH binaries with $m_t \sim 10^9 M_\odot$, as a result of enhanced accretion in the model, which was introduced to fit the GW background spectrum. This is compensated in our model by a larger number of mergers occurring at low redshift, allowing us to obtain a good match with the EPTA measurement.

Regarding LISA detections, our best estimate of a few hundred events per year (see Fig.~\ref{fig:lisa_rate}) is most consistent with the predictions of the ``Q3-nod'' model ($\sim$100–200 yr$^{-1}$) and the ``PopIII-d'' model ($\sim$50–100 yr$^{-1}$). In contrast, the predicted detection rates of $\sim$10 yr$^{-1}$ (``Loud'') and 10–20 yr$^{-1}$ (``Q3-d'') fall in the lower tail of our distribution, while the ``HS-nod-SN-high-accr'' model, with its prediction of $\sim$2000 yr$^{-1}$, lies in the upper tail. Detection rates are taken from Table 1 of~\cite{Barausse:2023yrx} for the ``Q3-nod'', ``Q3-d'', ``PopIII-d'', and ``HS-nod-SN-high-accr'' models, and from Table 1 of~\cite{2024A&A...686A.183I} for the "Loud" model.

This comparison demonstrates that, even in its current version, \texttt{POMPOCO} is flexible enough to describe a range of MBH populations -- a key goal of the model. To fully exploit the potential of LISA to distinguish between different formation mechanisms, however, it will be necessary  to explicitly include more than one formation channel in our model in the near future.

\section{Discussion and conclusions}\label{sec:conclusions}
In recent years, the number of observational probes into the population of MBHs in the Universe has been rapidly increasing, with even more expected in the near future. Preliminary results from JWST are already revealing intriguing discrepancies with previously accepted models, and it is anticipated that the upcoming observations will provide deeper insights. Additionally, PTA collaborations are on the verge of confirming the detection of the GW background in the nHz frequency band, generated by inspiralling MBH binaries. In the next decade, the launch of LISA will revolutionise our understanding of MBH formation and evolution by detecting a so far unobserved population of merging MBHs. To fully capitalise on these upcoming observations, it is essential to develop a framework that can comprehensively compare theoretical predictions with a wide variety of observational data.

To do this, we have developed our model \texttt{POMPOCO}: Parametrisation Of the Massive black hole POpulation for Comparison to Observations. It features 12 free parameters that are designed to effectively describe the formation and evolution of MBHs within DM halo merger trees. The computational efficiency of \texttt{POMPOCO}, compared to full cosmological simulations and SAMs, allow us to explore a wide range of parameter space and to identify the configurations that fit selected datasets best.

In this study, we used \texttt{POMPOCO} to demonstrate the consistency between the LF and the amplitude of the PTA GW background, particularly when allowing for an enhanced LF at high redshift, as suggested by extrapolating the results from \cite{Lacy:2015tha} and by the preliminary findings from JWST \citep{harikane2023jwstnirspeccensusbroadlineagns,Maiolino:2023bpi,Greene2024,Matthee2024,2025ApJ...986..165T}; see also \cite{2025ApJ...982..195S,Ellis:2024wdh,Padmanabhan:2024nvv} for a similar analysis. The authors found a consistency between PTA observations and JWST black hole candidates.)
To achieve this, we defined a likelihood function that accommodated the LF at different redshifts between the comprehensive fits from \cite{Shen:2020obl} and the fits to mid-IR observations from \cite{Lacy:2015tha}, and alongside the amplitude of the GW background inferred for the first frequency bin of the EPTA data. We then conducted a full Bayesian analysis. While our inferred GW spectrum clearlt agrees with the EPTA data, the median of our prediction lies slightly below theirs, which might indicate a lower background, as proposed by \cite{Goncharov:2024htb}. Although we focused on the case of circular binaries in this work, we hope to include the effect of eccentricity when fitting to PTA data in future studies.

We verified our results against the observed $M_*-M_{\rm BH}$ relation at $z =0$, for which we did not explicitly fit. The results agree very well with the fit by \cite{Greene2020}. 
Furthermore, our model is capable of reproducing even some of the MBHs that were observed outside of the main distribution. Many models struggle with this feature \citep{2021MNRAS.503.1940H}.

A key strength of \texttt{POMPOCO} lies in its ability to provide a posterior distribution for the model parameters when simultaneously fitting to the LF and the GW background. For instance, our Bayesian analysis showed that approximately $ 10\%$ of haloes with a mass $\gtrsim 10^7M_{\odot}$ at $z\sim 20$ should host an MBH seed, with characteristic masses ranging from $10^3$ and $5\times 10^6 M_{\odot}$. Additionally, we estimated that a fraction of MBHs undergoes super-Eddington accretion in burst episodes lasting a few tens of million years. At ``cosmic noon'', we predict an AGN fraction of $4-10\%$, and that the peak of accretion occurrs at $z\sim 2.5$. Finally, we found that MBHs with masses $\gtrsim 10^8M_{\odot}$ experience accretion suppression by $z\sim 2.5$, and that this  suppression extends to MBHs with masses $\gtrsim 6\times 10^6 M_{\odot}$ by $z=0$. These estimates are consistent with the findings of more detailed numerical simulations and observations. This suggests that \texttt{POMPOCO} captures the essential features that drive MBH formation and evolution.

Our model predictions for LISA are broad, but agree better with some scenarios than with others. In particular, they are more consistent with SAMs that assume that MBHs originate from heavy seeds and merge efficiently following galaxy mergers, such as the ``Q3-nod'' model~\citep{Klein:2015hvg} and ``HS-nod-SN-high-accr''~\citep{Barausse:2020mdt}. Discrepancies with the ``Q3-d''~\citep{Klein:2015hvg} and ``Loud''~\cite{2024A&A...686A.183I} models in the distribution of merger redshifts suggest the need of introducing a dependence of the post-galaxy-merger delay on the mass of the host galaxy. Overall, \texttt{POMPOCO} is flexible enough to describe a variety of scenarios for merging MBHs while remaining compatible with current observations. To leverage the full potential of LISA to distinguish between different formation scenarios, however, we plan to increase the flexibility of the seed distribution in our model.

Looking ahead, we aim to introduce several improvements to \texttt{POMPOCO}. In addition to incorporating eccentricity and other formation channels, we plan to model the spin evolution of MBHs, which is critically sensitive to the environment of merging MBH binaries \citep{Berti:2008af,Sesana:2014bea,2025PhRvD.111b3004S}. The inlcusion of eccentricity and spins will also allow us to correctly model the impact of kicks following MBH mergers, which we neglected in this work, and which could lower the LISA rates. In this work, we fixed the radiative efficiency to 0.1, whereas it is expected to depend on the accretion rate and on the spin of the BHs \citep[e.g.,][]{2008MNRAS.388.1011M,Madau:2014pta}. When spins and the suppression of the radiative efficiency at very low and very high Eddington ratios are included in our model, it will properly account for this. Moreover, we intend to adopt a more physically motivated accretion model, such as one based on the mass reservoir available for accretion. This might reduce the parameter space and prevent the formation of unrealistically large MBHs. Finally, we also plan to incorporate a dependence of the time delay between galaxy mergers and the formation of MBH binaries on the masses of the host galaxies. It will be crucial to address these issues for fully extracting the astrophysical information encoded in the LISA observations and realising the immense potential of the observatory \citep{Gair:2010bx,Klein:2015hvg,Bonetti:2018tpf,Toubiana:2021iuw,Fang:2022cso,Langen:2024ygz,2025PhRvD.111b3004S}.

\begin{acknowledgements}

We are thankful to S. Grunewald for his precious help in making \texttt{POMPOCO} fit to run on the Hypatia cluster, to L. Speri for providing us the EPTA data, and to M. Habouzit for fruitful discussion. A.T. is supported by ERC Starting Grant No.~945155--GWmining, Cariplo Foundation Grant No.~2021-0555, MUR PRIN Grant No.~2022-Z9X4XS, MUR Grant ``Progetto Dipartimenti di Eccellenza 2023-2027'' (BiCoQ), and the ICSC National Research Centre funded by NextGenerationEU. 
  L.S. acknowledges support from the UKRI guarantee funding (project no. EP/Y023706/1).   M.V. and S.B. acknowledge funding from the French National Research Agency (grant ANR-21-CE31-0026, project MBH\_waves). This work has received funding from the Centre National d’Etudes Spatiales.
   E.B. acknowledges support from the European Union’s H2020 ERC Consolidator Grant ``GRavity from Astrophysical to Microscopic Scales'' (Grant No. GRAMS-815673), the European Union’s Horizon  
ERC Synergy Grant ``Making Sense of the Unexpected in the Gravitational-Wave Sky'' (Grant No. GWSky-101167314), the PRIN 2022 grant ``GUVIRP - Gravity tests in the UltraViolet and InfraRed with Pulsar timing'', and the EU Horizon 2020 Research and Innovation Programme under the Marie Sklodowska-Curie Grant Agreement No. 101007855.  This work has been supported by the
Agenzia Spaziale Italiana (ASI), Project n. 2024-36-HH.0, ``Attività per
la fase B2/C della missione LISA''.

\end{acknowledgements}

\FloatBarrier
 \bibliographystyle{aa} 
\bibliography{Ref} 

\begin{appendix}

\section{Halo-black hole mass relation}\label{app:mhalo_mbh}

We show our $M_h-M_{\rm BH}$ relation at $z=0$ in Fig.~\ref{fig:mhalo_mbh}. Unlike in the $M_*-M_{\rm BH}$ relation, we observe no break at small MBH masses, reinforcing that the break in the latter originates from the $M_h-M_{*}$ relation of~\cite{2010ApJ...710..903M}.

The dotted line shows the extrapolation of the median at $M_h\lesssim 10^{14}M_{\odot}$ to larger halo values, obtained by performing a linear fit in the range $10^{13} M_{\odot} \leq M_h \leq 10^{15}M_{\odot}$. In order to estimate the impact of underestimating MBH masses at this end on the amplitude of the GW background (given by Eq.~\eqref{eq:hc_pta}), we have rescaled the masses of MBHs in haloes with $M_h\geq 10^{14}M_{\odot}$ by the ratio between the extrapolated median and our median. We found that this increases the background by $\sim 5\%$. This results in an increase in $\log_{10} h_c[1/(10 \ {\rm yr})] $ of $\sim 0.03  \ {\rm dex}$, which is ten times smaller than our statistical uncertainty. Moreover, the difference between the median reported by EPTA and ours is $\sim 0.2  \ {\rm dex}$, so this potential underestimation of the MBH masses in the most massive galaxies is likely not the reason behind the discrepancy. This is because the most massive galaxies, where the difference between the extrapolated median and ours is larger, are rare and contribute little to the background. In our model, their Press-Schechter weight suppresses their contribution.    

 \begin{figure}[h!]
\centering
\includegraphics[width=0.99\linewidth]{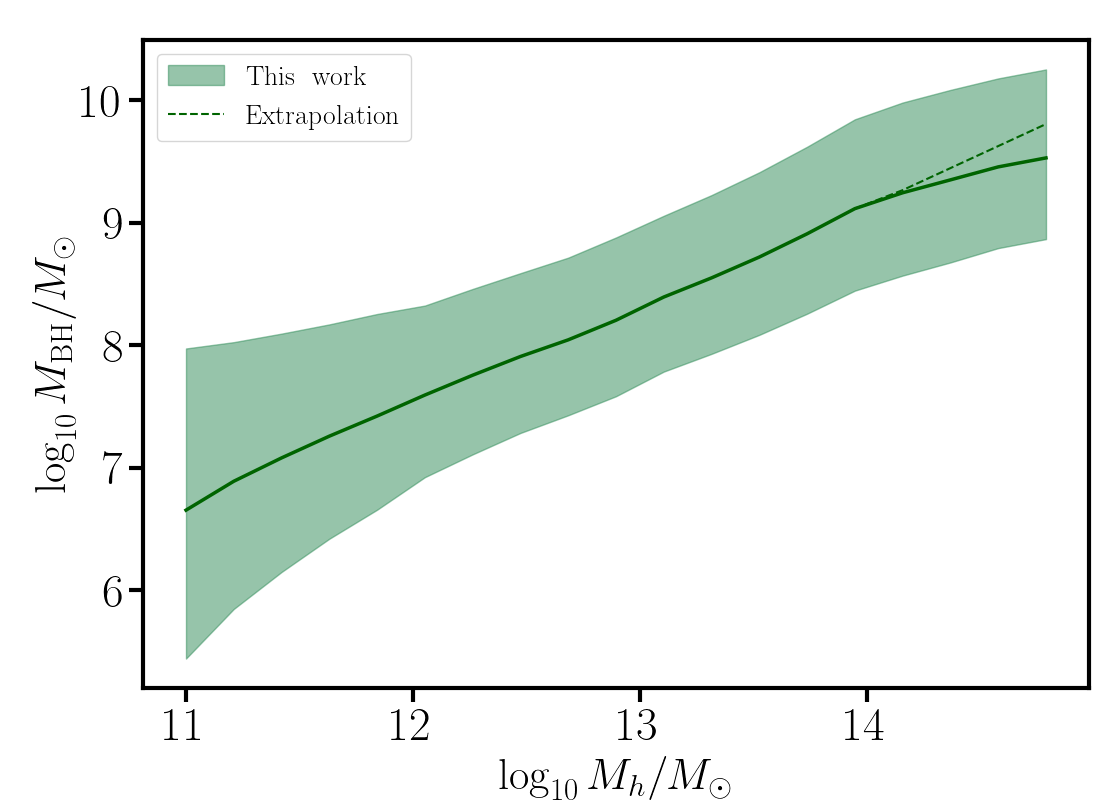}
   \caption{Relation between the halo mass and MBH mass at redshift $z=0$. }\label{fig:mhalo_mbh}
 \end{figure}

 \end{appendix}

\end{document}